# A Novel Approach to Personalized Personality Assessment with the Attachment-Caregiving Questionnaire (ACQ) – First Evidence in favor of AI-Oriented Inventory Designs

Marcantonio Gagliardi (*) – m.gagliardi@greenwich.ac.uk / 0000-0001-7275-6315

Marina Bonadeni (**) – Marina.Bonadeni@psypec.it / 0009-0004-5861-3421

Sara Billai (**) – Sara.Billai@psypec.it / 0009-0009-5805-9439

Gian Luca Marcialis (***) – marcialis@unica.it / 0000-0002-8719-9643

(*) University of Greenwich, London, UK, (**) Italian Professional Body of Psychologists & Psychotherapists, (***) University of Cagliari, Italy

Corresponding Author: Marcantonio Gagliardi (m.gagliardi@greenwich.ac.uk)

## Abstract

**Background.** Personality is a primary object of interest in clinical psychology and psychiatry. It is most often measured using questionnaires, which rely on Factor Analysis (FA) to identify essential domains corresponding to highly correlated questions/items that define a (sub)scale. This procedure implies the rigid assignment of each question to one scale – giving the item the same meaning regardless of how the respondent may interpret it – arguably affecting the assessment capability of the instrument.

**Methods.** To test this hypothesis, we use the Attachment-Caregiving Questionnaire (ACQ), a clinical and personality self-report that – through extra-scale information – allows the clinician to infer the possible different meanings subjects attribute to the items. Considering four psychotherapy patients, we compare the scoring of the ACQ provided by expert clinicians to the detailed information gained from therapy and the patients.

**Results.** Our analysis suggests that a question can be interpreted differently – receiving the same score for different (clinically relevant) reasons – potentially impacting personality assessment and clinical decision-making. Moreover, accounting for multiple interpretations requires a specific questionnaire design and a more advanced pattern recognition than FA – which Artificial Intelligence (AI) could provide.

**Conclusion.** Our results indicate that a meaning-sensitive, personalized read of a personality self-report can affect profiling and treatment. Since a machine learning model can mimic the interpretative performance of an expert clinician, our results also imply a novel, AI-oriented approach to inventory design, of which we envision the first implementation steps. More evidence is required to support these preliminary findings.

## 1. Introduction

Clinical psychology and psychiatry recognize the central role personal variables play in the development and maintenance of psychological vulnerabilities and mental disorders (APA, 2013; Beck et al., 2015; Wright & Hopwood, 2021). Despite being often used without specifications, the term '*personality*' refers to a complex concept underpinned by multiple features. Most abstractly, personality can be defined as the unique and (relatively) stable set of an individual's psychological characteristics – in terms of cognition, emotion, and behavior – usually described as traits (Beck et al., 2015; Corr & Matthews, 2009; Engler, 2013; Gagliardi, 2021; McCrae & Costa, 2003). These features are believed to be the product of innate biological factors and acquired knowledge, which confer personality uniqueness and stability.

Coherently, an individual is expected to manifest a recognizable personality – i.e., discernable cognitive, emotional, and behavioral expressions – across various contexts and situations. Nonetheless, subjects with similar traits may express them in different circumstances. Taking two *conscientious* individuals, for example, one could be very careful with their expenditures and neglect to tidy their home. The other, in contrast, may have the highest standards in keeping their living space in order but pay much less attention to money. This simple example illustrates how relevant the meaning subjects attribute to what they do can be. Therefore, to assess personality correctly, we should reasonably ensure that our instrument is capable of capturing meaning rather than behaviors in specific scenarios. Despite this potential issue, the standard validation procedure to develop a questionnaire – typically based on Factor Analysis (FA) or Principal Component Analysis (PCA)[1] (Dancey & Reidy, 2011; Gagliardi, 2022; Parsian & Dunning, 2009; Trochim et al., 2015) – aims to identify a few essential, relatively independent domains corresponding to highly correlated questions/items that identify a (sub)scale[2]. By doing that, it implies the attribution of a unique, fixed meaning to each item, corresponding to their rigid assignment to one and only one scale. In other words, an item having the same rating ('*I strongly agree*' on a Likert scale, for instance) from different respondents is given the same interpretation regardless of what the respondents meant.

In this work, we address the question of whether assigning unique interpretations to the items of a personality inventory can affect its capability to assess personality. We provide evidence that respondents can interpret an item differently, with a significant impact on scoring, suggesting an instrument should be designed to enable the attribution of multiple meanings. In other words, our data indicate that self-reports not designed to capture meaning – such as those based on FA – have an inherent limitation. And we should rely on more complex designs and methodologies to provide more flexible data interpretations.

Our study supports this proposition by considering psychological attachment – the evolutionarily preordained mechanism underpinning our innate motivation to seek care from a conspecific – which informs essential aspects of our personality through the early acquisition of data concerning the self, the caregiver, and the relationship (Bowlby, 1969/1982; Chopik et al., 2013; Gagliardi, 2021; Guidano, 1987, 1991; Karterud & Kongerslev, 2019; Levy et al., 2015; Mikulincer & Shaver, 2016; Noftle & Shaver, 2006; Young et al., 2019). We discuss the case studies of four patients who (1) completed the Attachment-Caregiving Questionnaire (ACQ) (Gagliardi, 2022) and then (2) started therapy, manifesting different clinical features over its course. The ACQ is an attachment-related personality inventory that includes default scales corresponding to the dimensions it aims to measure but also extra-scale sections with additional data. This design allows the scorer to interpret answers by cross-referencing information and generate flexible scales – i.e., assign an item to a non-default scale according to its interpretation. Indeed, the instrument is not intended to be validated following a standard procedure based on FA, which produces rigid groupings.

Treatment allowed us to test if patients giving similar ACQ answers regarding key clinical issues intended to convey similar or different meanings. In this respect, the ACQ – designed to detect what the respondent means by their answers – confirmed that flexible scales – provided by the possibility to interpret responses – can be essential to catch personal meanings and reach a more precise personality evaluation. Evidence suggests that a flexible, sensitive-to-meaning self-report can allow a human expert to improve personality assessment. Since an Artificial Intelligence (AI) model – such as a neural network – could learn how to mimic this interpretative performance, it also implies a novel, AI-oriented approach to inventory design, of which we will outline the possible first implementation steps.

[1] Although FA and PCA differ, their differences are usually irrelevant, and researchers refer to PCA as a kind of FA. For simplicity, we follow the same practice.

[2] A questionnaire itself can be considered a scale, possibly consisting of multiple sub-scales. Here, we refer to the sub-scales as scales.

We proceed as follows. We first set the background by presenting a representative, well-established model of personality assessment (section 1.1). Then, focusing on attachment and personal meaning, we review several classical self-reports measuring these constructs (sections 1.2, 1.3) and describe the ACQ, highlighting its distinctive features (section 1.4). Finally, we illustrate our four clinical cases (section 2), present the ACQ results (section 3), and discuss them (section 4).

### 1.1 The Big Five and the standard assessment of personality

Despite being an intuitive concept, personality is hard to define, and many models have been proposed (Engler, 2013; Friedman & Schustack, 2015). Among them, the most accredited in research is probably the Five Factor Model (FFM) (McCrae & Costa, 2003) – more commonly known as the Big Five Model – identifying the five personality dimensions (or traits): *Openness*, *Conscientiousness*, *Extraversion*, *Agreeableness*, and *Neuroticism* (OCEAN).

The FFM derives from a standard FA-informed procedure, which identifies highly correlated groups of items constituting a scale (Dancey & Reidy, 2011; Gagliardi, 2022; Parsian & Dunning, 2009; Trochim et al., 2015). In this case: (1) Self-descriptive items were formulated using common language descriptors of personality characteristics. (2) A large sample rated such self-referred items on a Likert scale. And finally, (3) FA was used to extract the underlying dimensions. The best characterization of personality turned out to be provided by five groups of descriptors – forming the Big Five. Multiple self-reports measure these traits – such as the NEO Personality Inventory (NEO-PI) and its revised version (NEO-PI-R) (Costa & McCrae, 2009) and the Big Five Inventory (BFI) (John & Srivastava, 1999).

The standard factorial method leads to personality inventories consisting of multiple scales, each corresponding to a dimension – the five OCEAN traits in the case of the NEO or BFI, for example. These scales are rigid – with items bounded to them – which, in terms of interpretation, corresponds to assuming every subject *reads* each item in the same way. With the BFI, for instance, 'being talkative' is always associated with extraversion, and 'worrying a lot' is tied to neuroticism. Possible different attributions of meaning – i.e., different psychological experiences corresponding to the same descriptor – are not taken into account. However, given the potential relevance of alternative experiences (e.g., 'being talkative' out of openness and not extraversion or 'worrying a lot' out of conscientiousness and not neuroticism), we suggest this omission to be a significant limitation when assessing personality (limitation 1, L1).

It is worth noting that another inherent limitation of the FFM is relying only on common descriptors to identify personality dimensions – since they do not reflect the whole range of psychologically relevant experiences. Some have never been given a name – like those states necessitating long explanations to be communicated or for which only technical terms are available (a particular sense of guilt or a dissociative manifestation, for instance). As a result, assessing the Big Five can only tap into the vocabulary-emergent psychological experiences (limitation 2, L2).

This work addresses L1 through four case studies, relying on a non-standard, attachment-related personality inventory – the ACQ. Therefore, we next review the classical, FA-based self-reports that inspired its development (sections 1.2, 1.3) and then illustrate the ACQ and its features (section 1.4).

## 1.2 Attachment questionnaires

Attachment has a dimensional nature, and classical research identified the three dimensions of *disorganization*, *avoidance*, and *ambivalence*[3] (Fraley et al., 2015; Paetzold et al., 2015). While disorganization refers to traumatic experiences and is usually considered a delicate case of specific clinical interest, the other two dimensions underpin our usual experience. Avoidance concerns one's tendency to seek an emotional connection with the caregiver, and ambivalence pertains to the degree to which one requires the caregiver's availability. These two are typically measured using self-reports referring to attachment in romantic relationships (Barone & Del Corno, 2006; Ravitz et al., 2010). Below, we review three of the most used of these instruments.

- **The Adult Attachment Scale (AAS).** The AAS (Collins & Read, 1990) and revised AAS (RAAS) (Collins, 1996) assess the two dimensions of avoidance and ambivalence through 18 items – 12 items for the first dimension and 6 for the second one (Brennan et al., 1998; Collins et al., 2006). The items were created by referring to existing literature, particularly the attachment style descriptions from Hazan and Shaver (1987). Respondents are prompted to reflect on their romantic relationships and the related emotional experiences.
- **The Adult Attachment Questionnaire (AAQ).** The AAQ (Simpson et al., 1992; Simpson et al., 1996) evaluates the attachment avoidant and ambivalent dimensions using 17 items – 8 for the former and 9 for the latter. Again, the attachment style descriptions from Hazan and Shaver (1987) served as a primary inspiration for item development. Subjects are asked to indicate how they typically feel toward romantic partners.
- **The Experiences in Close Relationships (ECR).** The ECR (Brennan et al., 1998) and ECR revised (ECR-R) (Fraley et al., 2000) consist of 36 items measuring avoidance and ambivalence (18 items for each dimension) – which were derived from an initial pool created by drawing from the available attachment literature. As with the AAS and AAQ, subjects are invited to refer to how they generally experience relationships.

Avoidants feel uncomfortable being close and dependent on the other. They tend to shun emotional intimacy and suppress emotional needs – correspondingly, they deactivate attachment (Shaver & Mikulincer, 2002). Ambivalents fear the other to be unavailable when needed. They tend to be concerned about being unattended and left – correspondingly, they hyper-activate attachment (Shaver & Mikulincer, 2002). The secure attachment style emerges when both avoidance and ambivalence scores are low.

In all cases, researchers applied FA to the collected data to find the underlying item structure given by their correlations. They identified how items clustered together into meaningful factors/dimensions – constituting the questionnaire scales. These latent factors – i.e., the associated items – represented avoidance and ambivalence. In the ECR case, the avoidant and ambivalent 18-item scales were derived by applying FA to 323 questions corresponding to 60 attachment constructs.

## 1.3 Personal meaning questionnaires

Following Bowlby's (1969/1982) perspective – according to which early caregiving critically forges personality, Guidano (1987, 1991) identified four traits closely related to attachment experiences that he termed phobic, depressive, eating disorder, and obsessive personal meaning organization (PMO). These traits derive their names from specific mental disorders, representing their most probable extreme manifestations. Their central features can be described as follows:

**(1) Phobic (P) Organization.** The P-organized is particularly sensitive to the dichotomy of '*being close to the reference figure to receive protection*' and '*being free to explore*'. The need for protection corresponds to perceiving the world as dangerous and being vulnerable to separation anxiety. As a result, the P-organized tends to focus on their physiological expressions as a signal of danger to their health, overlooking alternative explanations, especially relational ones.

**(2) Depressive (D) Organization.** The D-organized is particularly sensitive to having their value recognized by their reference figure and, therefore, oriented toward reaching some evident achievements. Perceiving the lack of recognition causes an underlying sense of defeat and irremediable loss, corresponding to inherent solitude. As a result, the D-organized tends to attribute the meaning of loss to life events, particularly concerning their relationships, and systematically rely on themselves.

**(3) Eating Disorder (ED) Organization.** The ED-organized is particularly sensitive to their reference figure's confirmation in order to define themselves, primarily their sensations and emotions. The sense of uncertainty about oneself – rooted in being unsure about the interpretation of one's somatic state – leads to relying on external references and being approved by others to define one's own thoughts and feelings. As a result, the ED-organized tends to follow expectations and comply with social standards.

**(4) Obsessive (O) Organization.** The O-organized is particularly sensitive to appearing as a good person – primarily to their reference figure – by strictly abiding by a given set of rules. Respecting this code – which distinguishes between right and wrong – determines one's intrinsic nature and is, therefore, essential. Rules can be more or less abstract, from general principles to specific ways to operate in particular domains. As a result, the O-organized tends to control their actions with respect to the domains involved by their rules.

[3] In the attachment literature, ambivalence is often referred to as *anxiety.* But, given the clinical nature of our work, this term may improperly suggest a clinical condition rather than an attachment style.

Relying on Guidano's findings, researchers developed two inventories to measure these traits, which we review below.

- **The Personal-Meaning-Questionnaire (PMQ).** The PMQ (Picardi & Mannino, 2001; Picardi et al., 2003) assesses the four PMOs through 68 items – 17 for each PMO. Item formulation was inspired by the accurate trait descriptions from the literature (Guidano, 1987, 1991). Subjects are asked to rate each statement according to how much it applies to them and provided with a corresponding personality profile over the four dimensions.
- **The Mini Questionnaire of Personal Organization (MQPO).** The MQPO (Nardi et al., 2012) measures the four Guidanian PMOs according to the theoretical evolution developed by Nardi (Nardi & Bellantuono, 2008), using 5 items for each trait for a total of 20 items. As usual, the items were created starting from literature descriptions. Subjects rate them for their compliance with personal characteristics and obtain their four-dimensional personality profile.

As with the other reviewed questionnaires, the PMQ and MQPO were developed using FA to create four scales, each measuring a specific dimension. In this case, statements clustered together refer to a particular PMO. Central to our argument is that all the above self-reports resulted in the creation of rigid scales – where items have a unique meaning, regardless of what the subject actually means by rating them.

### 1.4 The Attachment-Caregiving Questionnaire

Our study relies on the Attachment-Caregiving Questionnaire (ACQ) (Gagliardi, 2022), a 394-item personality and clinical inventory, whose development started from analyzing the extant attachment and personal meaning self-reports (cf. 1.2, 1.3) but followed a unifying theoretical framework. It refers to the Attachment-Personality Theory (APT) (Gagliardi, 2021, 2024), which proposes an extension of the 3D classical attachment theory by introducing four additional dimensions – *phobicity*, *depressivity*, *somaticity*, and *obsessivity*. These dimensions have an evolutionary origin, and their manifestations largely overlap with the Guidanian personal meaning organizations (in the order followed above). This extended dimensional framework provides a (more) comprehensive range of attachment experiences and related acquisition of personality features, including possible vulnerabilities to mental disorders.

The ACQ was designed to measure these seven attachment dimensions in adulthood with two main goals: (1) Personality profiling. (2) Clinical assessment. (1) Given the relevant role played by attachment in structuring personality (Bowlby, 1969/1982; Chopik et al., 2013; Gagliardi, 2021; Guidano, 1987, 1991; Karterud & Kongerslev, 2019; Levy et al., 2015; Mikulincer & Shaver, 2016; Noftle & Shaver, 2006; Young et al., 2019), its measure provides an attachment-related personality evaluation. Therefore, the ACQ works as a personality inventory that generates a seven-dimensional profile (each dimension represents an attachment-related trait). (2) On the other hand, given the link between attachment and psychopathology (DeKlyen & Greenberg, 2016; Gagliardi, 2021; Guidano, 1991; Liotti, 2011; Lyons-Ruth & Jacobvitz, 2016; Nardi & Bellantuono, 2008; Stovall-McClough & Dozier, 2016), including the most common psychological conditions – such as anxiety, mood, eating, and obsessive disorders – measuring attachment dimensions is clinically relevant. Therefore, the ACQ also works as a clinical test that evaluates the patient's attachment-related vulnerabilities to mental disorders.

Following the usual practice, the ACQ was realized by drawing on the extant literature but now allowing the scorer to interpret items, aiming to match what the respondents mean by their answers. The questionnaire includes 'default scales' directly linked to specific dimensions – 16 items for disorganization, 18 for avoidance, 15 for ambivalence, and 19 for phobicity, depressivity, somaticity, and obsessivity. However, these scales are flexible since the scorer can assign their items to a different scale depending on their interpretation. For example, the scorer could 'move' a default ambivalent item to the depressive scale (if they reckon the respondent gave that item a depressive meaning). This feature is enabled by including extra-scale information that allows the scorer to interpret answers by cross-referencing data. Overall, the ACQ consists of three sections: (1) contextual data, (2) current attachment state (default scales), and (3) childhood caregiving experiences (Table 1).

| ACQ Section | | Subsection | | Items | Description |
|---|---|---|---|---|---|
| 1 | **Contextual Data** | A | Personal Information | 23 | Data on the subject's context – current life environment and clinically relevant information. |
| | | B | General Condition | 20 | |
| | | C | Specific Issues | 17 | |
| 2 | **Attachment** | A | Introduction | 3 | Current attachment state. Default scales for seven attachment-related personality traits. |
| | | B | Attachment | 125 | |
| 3 | **Caregiving** | A | Introduction | 1 | Childhood caregiving experiences with the two most relevant attachment figures and in the family in general. |
| | | B | Family | 17 | |
| | | C | Introduction | 4 | |
| | | D | Maternal Figure | 83 | |
| | | E | Introduction | 4 | |
| | | F | Paternal Figure | 83 | |
| | | G | Additional Information | 14 | |
| | | | | 394 | |

**Table 1.** ACQ structure. The ACQ is structured into three sections – (1) Contextual Data, (2) Attachment (default scales), and (3) Caregiving – gathering information on the subject's context, current attachment state, and childhood caregiving experiences, respectively. The relevance and range of these data allow for interpreting the default scale items by cross-referencing information.

In conclusion, the ACQ design allows a scorer (e.g., a clinician) to build a comprehensive picture of the subject's state and attribute meaning to the items accordingly. In contrast to standard self-reports, this questionnaire is not intended for classical FA-informed validation. Indeed, its objective of implementing flexible scales would be inconsistent with such a validating procedure – which implies rigid item allocations. However, an AI model – such as a neural network – can be trained to mimic the scorer's performance, thereby also providing a non-human validation of the instrument (if an AI model can learn a pattern, that proves there is a pattern to learn).

We now describe how we used the ACQ and external information to test if scale flexibility is relevant to personality assessment.

## 2. Methods

This study relies on four clinical cases concerning patients who completed the ACQ before starting therapy. They received treatment from one of the first three authors and gave informed consent to use the questionnaire data and therapy material for research purposes. The ACQ received ethics approval first from the University of Sheffield (Ref. 032300) and then from the University of Greenwich (Ref. 21.5.7.14). A copy is included as supplementary material.

We assessed these patients' personality in relation to specific attachment dimensions using the ACQ and throughout treatment (Table 4). The second and the fourth patients gave self-report responses that partially overlapped, respectively, with those provided by the first and third patients on their prevalent dimension – posing the problem of how to interpret such answers and score the questionnaire. A corresponding clarification of personal meanings unfolded throughout sessions as the result of standard treatment. We present below a summary of the cases (cf. 2.1) and the procedure we followed to analyze them (cf. 2.2).

### 2.1 Participants

The participants in this study were four patients who asked for assistance in reducing the distress arising from their life circumstances (as described below). All were treated for 18 months with weekly sessions – long enough to collect detailed clinical information and formulate an accurate case. What follows is a synopsis of their clinical cases at the beginning of therapy (names are fictional).

### 2.1.1 Case Study 1 – Harry

Harry is a 40-year-old nurse who loves his job, sports, and traveling. He is a friendly man and extremely helpful to the people he takes care of, as well as to his friends and his brother. At the moment of starting therapy, Harry is not in a relationship, but he sought help to address the conflictual dynamics with his ex-partner. Since Harry was a child, his parents' arguments and neglect of his needs have triggered great anger in him. Reflecting on these circumstances, Harry used to say to himself that one should choose their partner very carefully. However, looking at his last relationship, he reports many instances of feeling sidelined by his ex and unimportant. Harry used to protest, but her job, friends, and many interests seemed always to come before him. Not even when he asked her to marry him and create their own family could he catch her attention. Despite being aware of his dissatisfaction, Harry kept following the same patterns until she finally left him for another man.

### 2.1.2 Case Study 2 – Erika

Erika – a 28-year-old nightclub bartender who lives with her mother – has been feeling depressed for over six months, experiencing a loss of hope in the future, her abilities, and the people around her, from whom she is constantly disappointed. She believes she has made poor choices that have led her to her current failure. When Erika was eight, her parents separated, and her father left. A few years later, she reached out to him, who became a part of her life again. Now, her bond with her father seems stronger. Erika admires him, and he sometimes reciprocates by praising her skills and perfectionism. However, she views herself as flawed, incapable of maintaining friendships, yearning for recognition that never comes, and appearing as a loser in comparison with others. Working overnight, Erika spends her free daylight hours in her room, haunted by a profound sense of loneliness. She interacts with only a few people, including her parents and the man with whom she has an intermittent relationship. Her romantic relationships do not last long.

### 2.1.3 Case Study 3 – Jordan

Jordan is 24 years old as he approaches the conclusion of his university studies and works part-time at an accountant's office. Lately, he has seen an increase in the obsessions that were already present since adolescence, during which he was diagnosed with Obsessive-Compulsive Disorder (OCD). Jordan recounts a childhood poor of parental affection and characterized by precise and rigid family rules. He was particularly struck by an episode at the age of 16 when – after informing his mother he had kissed the young girl he was dating – the mother blamed him for committing a grave act that required not only confession but also to undertake a path of 'true' redemption. She accused him of severely hurting her by not preparing her to face the fact he was growing up and forbade him to keep seeing the girl. During the last months, Jordan has returned home in the evening and incessantly dwelled on what happened at work. He experiences intrusive thoughts about not completing all tasks correctly and feels compelled to retrace in his mind every step taken throughout the workday. He is also distressed at the thought his inattentiveness might have harmed a colleague.

### 2.1.4 Case Study 4 – Beth

Beth is a data scientist aged 35. She sought psychological support for managing a period of particular stress. Her boyfriend was away most of the time for work. She was working on an important project under a lot of pressure from her boss. And her parents were becoming particularly demanding, asking for her support with family matters. One year through therapy, it started to become clear that Beth's childhood experiences had a relevant impact on how she tended to feel and think. In particular, she reported how her mother used to push her toward being a professional dancer – an activity everybody would admire. She has present in her mind the endless hours of training with her team and how disappointed the mother was when she did not perform as expected. She gradually managed to distance herself from professional dancing and to pursue a career as a data scientist, landing in a top software company. Nonetheless, she realizes how hard it can be for her to be herself without disappointing the expectations of others and take life easier. Even with her friends, she always tries to be the ‘perfect’ one and pays much attention to every comment on her.

## 2.2 Procedure

The study aimed to test the ACQ’s potential to improve personality assessment by interpreting unclear answers (i.e., those potentially not belonging to their default scale because the respondent might have read them differently). Patients completed the ACQ, but – to ensure treatment was not influenced by what they had reported – the outcomes remained undisclosed to them and their therapist. After 18 months, the knowledge gained in therapy was used to test the ACQ interpretation. More specifically, the study followed the four-stage design outlined below (S1-S4) (Figure 1).

**(S1) Start of therapy**. Recruitment took place during a two-week time frame. The initial participants were the seven patients who started therapy with one of the first three authors in that time frame and agreed to participate. Based on the authors’ clinical experience, an 18-month standard treatment (t-period) was estimated to be necessary (and reasonably sufficient) to explore the attachment-related meanings most relevant to a patient. As a result, this condition was considered essential for inclusion in the post-treatment stages of the study.

**(S2a) ACQ blind scoring**. Starting therapy, the seven participants completed the ACQ, generating a personality profile on the seven attachment dimensions of disorganization, avoidance, ambivalence, phobicity, depressivity, somaticity, and obsessivity (Gagliardi, 2021, 2024). Each self-report was scored by the two authors who did not treat the respondent. The scoring was carried out independently (97% inter-rater agreement) and later discussed to agree on a final ‘*blind*’ profile (i.e., not informed by clinical knowledge of the patient). Therapists had no access to their patients’ questionnaires for the first 18 months.

**(S2b) 18-month therapy (t-period)**. Therapists conducted therapy following cognitive-evolutionary principles (Gagliardi, 2021; Ivaldi, 2016; Liotti, 2011; Liotti et al., 2017) with a focus on the patient’s relationships, cognitive structures, and motivational dynamics, using the therapeutic relationship as an active therapeutic tool (Baier et al., 2020; Liotti & Monticelli, 2014; Slade & Holmes, 2019). Attention was given to the attachment-related meanings most relevant to the patient, aiming to clarify them in the light of critical events. After the t-period, four patients had attended regularly and could enter the following stages of the study. The other three were excluded – one had completed treatment after a little over a year, and two had attended inconsistently.

**(S3) ACQ scoring test 1**. At this point, the therapists provided a detailed written profile of their patients guided by their session notes – without any knowledge of the patients' ACQs. The authors then used this material to test the initial ACQ blind interpretation against the information gained in treatment, focusing on specific answers underpinned by possible alternative meanings.

**(S4) ACQ scoring test 2**. Finally, therapists explicitly invited the four patients to elaborate in writing on their Am1-Am4 and Ob1-Ob4 answers to test the ACQ interpretation against patients’ explanations. Patients all received the same hints for each item – i.e., “What motivated you to answer with this score?”, “What were your thoughts?”, “What were your feelings?”, and “Had you one or more specific episodes in mind?”.

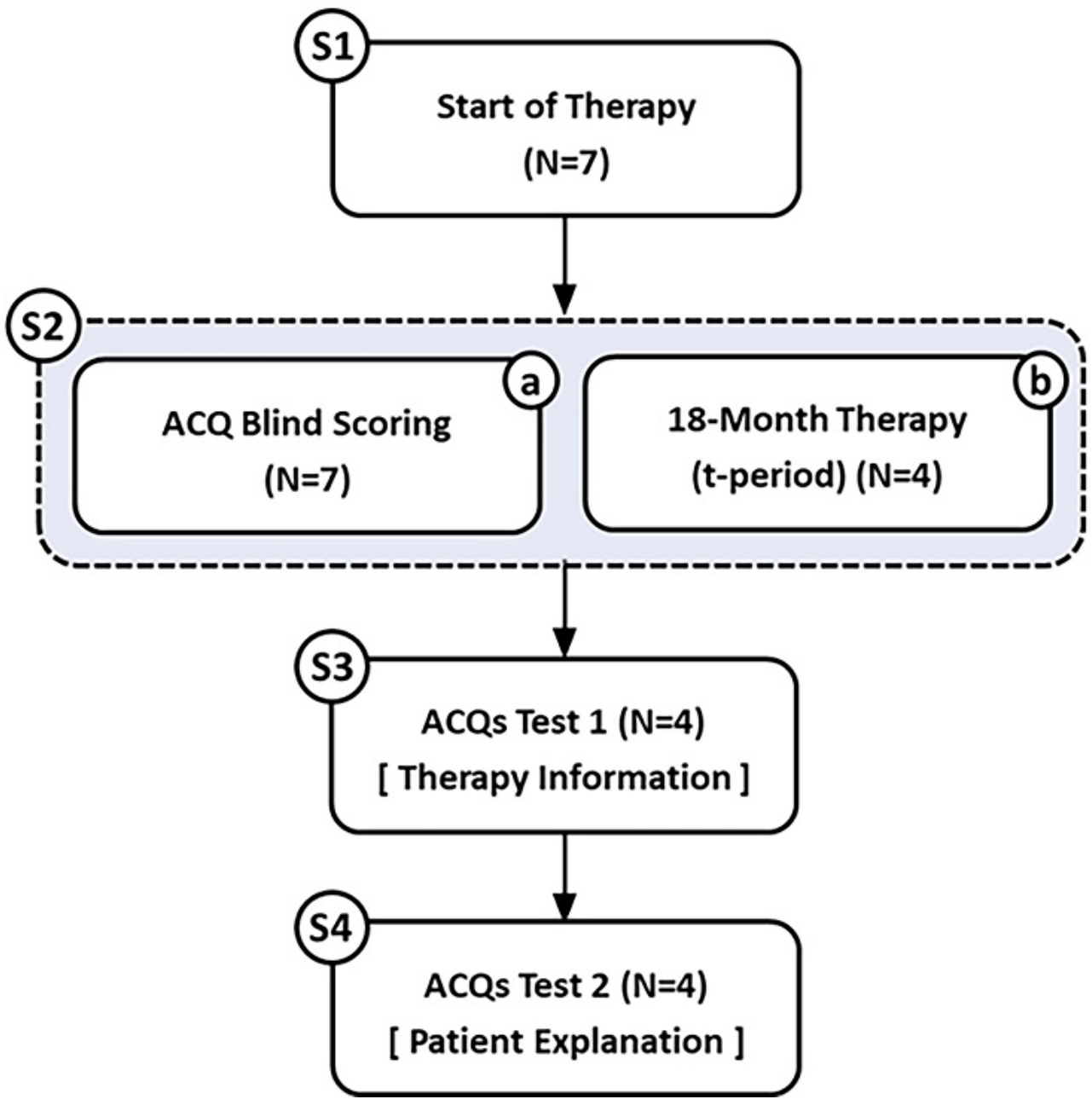

**Figure 1.** Study design. The study was designed in four stages (S1-S4). (S1) It started with seven participants consisting of the patients who began therapy in the same two-week period and agreed to participate. (S2) These patients completed the ACQ – whose scores were not disclosed to the therapists (S2a). After 18 months (t-period), four patients were still in treatment and continued the study (S2b). Their ACQs presented two primary interpretation issues – on the ambivalent and obsessive (default) scales – which were tested against (S3) therapy information and (S4) patients' explanations of their answers.

## 3. Results

We report the results of our analysis following the four stages outlined above.

**(S1) Start of therapy**. Seven patients started therapy, but only four – Harry, Erika, Jordan, and Beth – attended 18 months of treatment regularly and completed the study.

**(S2a) ACQ blind scoring**. On the ACQ, Harry and Erika reported high scores on the ambivalent scale – significant concerns about rejection and abandonment. Jordan and Beth scored high on the obsessive one – a marked propensity to abide by given rules and worry about doing the right thing. Such features are recognized to be typical of ambivalence and obsessivity, respectively (Collins, 1996; Fraley et al., 2000; Gagliardi, 2021; Nardi et al., 2012). However, the scorers deemed two patients gave several items of these scales a non-default meaning (i.e., despite belonging to the ambivalent/obsessive scale, they were read by the respondent as an item of another scale). This was the case for items Am1-Am4 and Ob1-Ob4 (Table 2), whose ratings by the four patients are reported below (Table 3).

| (a) | Four items from the ACQ ambivalent scale |
|---|---|
| Am1 | In a relationship, the idea of being left by my partner hardly enters my mind |
| Am2 | In a relationship, I'm confident my partner would never leave me |
| Am3 | In a relationship, I think of what I'd do if my partner left me |
| Am4 | In a relationship, I wonder whether my partner really cares about me |

| (b) | Four items from the ACQ obsessive scale |
|---|---|
| Ob1 | Not respecting my rules would be unacceptable to me |
| Ob2 | Moral issues – what is right or wrong – are at the heart of my thoughts |
| Ob3 | Always doing the right thing is essential |
| Ob4 | The slightest doubt that I have done something wrong can make me feel terrible anguish |

**Table 2.** Eight ACQ items. (a) The upper part of the table shows four items from the ACQ ambivalent scale – they concern worrying about rejection and abandonment. (b) The lower part of the table presents four items from the ACQ obsessive scale – they concern the tendency to follow rules and the worry of doing the right thing.

| (a) | (1) Harry | (2) Erika |
|---|---|---|
| Am1 | 8 | 10 |
| Am2 | 9 | 7 |
| Am3 | 8 | 8 |
| Am4 | 10 | 9 |
| Mean | 8.75 | 8.50 |

| (b) | (3) Jordan | (4) Beth |
|---|---|---|
| Ob1 | 9 | 10 |
| Ob2 | 9 | 9 |
| Ob3 | 8 | 9 |
| Ob4 | 10 | 7 |
| Mean | 9.00 | 8.75 |

**Table 3.** Sample ACQ scores. (a) The first two patients – Harry and Erika – scored similarly on items Am1-Am4, characterizing ambivalence (Table 2) (Am1 and Am2 indicate reversed scores). (b) The other two – Jordan and Beth – gave similar ratings to items Ob1-Ob4, inherent to obsessivity.

Harry’s and Erika’s means were comparably high on the ACQ ambivalent items (8.75 vs. 8.5), and similarly, Jordan’s and Beth’s means on the ACQ obsessive items (9.00 vs. 8.75). As a result, these items alone would make it difficult to detect a possible difference between the patients’ ambivalence and obsessivity levels. Nonetheless, a non-default meaning attribution – in this case, not ambivalent or obsessive – was enabled by the possibility given by the ACQ to consider information from other parts of the questionnaire.

First, the two pairs of patients differed on other scales. In particular, Erika’s predominant scale was depressivity, and Beth’s was somaticity. Moreover, other items suggested possible non-standard interpretations of Am1-Am4 from Erika and Ob1-Ob4 from Beth. Erika was previously diagnosed with depression and reported depressive experiences during childhood in relation to her mother (e.g., not having the chance to spend time with the mother, missing the mother, and longing to be recognized by the mother as valuable). Beth reported feeling significantly under pressure due to others’ expectations of her and the experience of a typical somatic childhood with both parents (e.g., high admiration for the parents and marked dependence on the parents’ approval). Finally, Erika’s and Beth’s life histories emerging from their questionnaires were inconsistent with the high ambivalent and obsessive manifestations suggested by the default scales.

On these grounds, while Harry’s and Jordan’s ACQ scores appeared driven by their ambivalence and obsessivity, respectively, Erika’s ambivalent answers were recognized to be underpinned by her depressivity and Beth’s obsessive answers by her somaticity (Table 4). In other words, the ACQ included enough data to deem it more probable that Erika and Beth attributed Am1-Am4 and Ob1-Ob4 depressive and somatic meanings, respectively.

| (a) | Case Study | Prevalent Dimension | Additional ACQ information used to score Am1-Am4 |
|---|---|---|---|
| 1 | Harry | Ambivalence | None [ambivalence as prevalent dimension] |
| 2 | Erika | Depressivity | (1) Depressivity scores [depressivity as prevalent dimension]<br>(2) Previous depression diagnosis<br>(3) Childhood depressive experiences with mother<br>(4) Life history inconsistent with prevalent ambivalence |

| (b) | Case Study | Prevalent Dimension | Additional ACQ information used to score Ob1-Ob4 |
|---|---|---|---|
| 3 | Jordan | Obsessivity | None [obsessivity as prevalent dimension] |
| 4 | Beth | Somaticity | (1) Somaticity scores [somaticity as prevalent dimension]<br>(2) Feeling significantly under pressure due to others' expectations<br>(3) Childhood somatic experiences with parents<br>(4) Life history inconsistent with prevalent obsessivity |

**Table 4.** ACQ interpretation of ambivalent and obsessive items. (a) Harry and Erika gave similar ratings of Am1-Am4. However, additional ACQ information allowed us to deem Erika's answers on these items as having a depressive meaning. (b) Similarly, Jordan's and Beth's ratings of Ob1-Ob4 were comparable. Nonetheless, additional ACQ information allowed us to reckon Beth's answers on these items as having a somatic meaning.

**(S2b) 18-month therapy (t-period)**. The 18 months of therapy planned for the study allowed the therapists to formulate a reliable evaluation of the patients' attachment-related personality traits. In particular, two clinically relevant themes related to Am1-Am4 and Ob1-Ob4 emerged in therapy: (T1) A sense of loneliness – expressed by worries about being rejected and abandoned – crucial to the first two patients (Harry and Erika). (T2) The tendency to follow (some) rules, characterizing the other two (Jordan and Beth).

(1) Harry's ambivalent attitude was clear early on in treatment. His anger and worry about being disregarded and left alone were central to the interpretation of his life events. The therapist recognized a prevalent ambivalent trait characterized by marked protest toward his reference figures. (2) On the other hand, Erika had sought help for her depressed mood. She seemed to have never overcome the loss of her father, although she could reconnect with him after some time. Her sense of solitude and the impossibility of feeling valued by her significant others dominated her experience. (3) Jordan had a history of OCD. He focused on following the rules that made him feel to be a good person. Obsessivity was clearly his primary trait. (4) In contrast, Beth focused on how she appeared to others and found it extremely hard to express her true self. She felt oppressed by others' expectations, especially from her mother, and following rigid rules was her way to feel adequate and accepted.

The elaboration of T1 and T2 throughout therapy clarified the actual experiences and clinical nature related to them, thereby allowing for testing the interpretation of the corresponding questionnaire items.

**(S3) ACQ scoring test 1**. The ACQ blind interpretation of the ambivalent and obsessive items – in particular, Ab1-Ab4 and Ob1-Ob4 – was first tested against the information gathered by the therapists throughout the t-period and reported in the detailed profiles therapists had prepared. (1) Erika showed a depressive profile with a profound sense of loneliness consistent with high scores on the ambivalent items touching on rejection and abandonment. Moreover, she never showed marked signs of ambivalence – such as anger and protest for having her needs unmet. (2) Similarly, Beth was soon identified as predominantly somatic. She was strongly dependent on reference figures for approval and tended to follow rules to feel aligned with others' expectations. Beth never showed signs of obsessivity – such as feeling compelled to do some actions to avoid terrible consequences. What she tended to term a sense of guilt was informed by failing at being included rather than causing some harm, as obsessivity would suggest. Overall, the profiles of these two patients that emerged in therapy supported the non-default interpretations of their ACQ ambivalent and obsessive scores – as depressive and somatic, respectively.

**(S4) ACQ scoring test 2**. Finally, the examination of patients' writings on items Am1-Am4 and Ob1-Ob4 confirmed the profiles therapists had formulated throughout treatment, pointing to Harry's, Erika's, Jordan's, and Beth's answers as being driven by a predominant ambivalent, depressive, obsessive, and somatic meaning, respectively. No information emerged that could reasonably suggest a different interpretation.

## 4. Discussion

This study investigates the potential of a questionnaire to assess personality. We put forward the hypothesis that self-reports developed through classical statistical methods – such as Factor Analysis (FA) – are intrinsically limited. They identify items that statistically cluster together but cannot capture what motivates answers. To overcome this limitation, we suggest adopting more complex designs and methodologies that allow for the interpretation of responses by cross-referencing data from various sections and scales. This pattern recognition ability belongs to expert humans (clinicians, for instance), but we know it can be reproduced by an Artificial Intelligence (AI) model – like a neural network – adequately trained (Bishop, 2006; Dougherty, 2012).

To test our hypothesis, we examined the cases of four patients who completed the Attachment-Caregiving Questionnaire (ACQ) (Gagliardi, 2022) and underwent psychotherapy for 18 months. The ACQ is an attachment-related personality inventory (for the link attachment-personality, e.g., Chopik et al., 2013; Gagliardi, 2021; Karterud & Kongerslev, 2019; Levy et al., 2015; Young et al., 2019) that allows the therapist to build a picture of the respondent's life context (ACQ first section), current state of mind (ACQ second section), and childhood caregiving experiences (ACQ third section). This information can be used to interpret ambiguous answers – i.e., answers that may not have been interpreted by the respondent according to their default scale.

Soon after deciding to start therapy and participate in the study **(S1)**, patients completed the ACQ. The questionnaires were blind-scored **(S2a)** and kept undisclosed for the first 18 months of treatment (t-period) **(S2b)**. Finally, the ACQ interpretations were tested against the knowledge acquired throughout therapy **(S3)** and the additional information provided by the patients **(S4)**.

**(S2a)** Despite the efforts of a questionnaire designer, respondents can interpret items in multiple ways. The statement "*The relationship with my mother was affectionate*", for example, can have radically different meanings for two different individuals – from "*We sometimes went to the park together*" to "*My mom cuddled and reassured me when I was down*". Nonetheless, these two individuals can rate that statement similarly on the questionnaire. An unexpected meaning attribution cannot be detected by usual, rigid-scale instruments but can be spotted if more comprehensive information is provided and data can be cross-referenced. The deriving scale flexibility is what the ACQ enables. The clinician can interrelate multiple pieces of information to make sense of their patient's ambiguous responses. When located in a broader context, items assume a specific meaning and contribute to building an individual story.

From the ACQ, the scorers could build a narrative of each patient's experience consistent with their life context, current state of mind, and childhood caregiving experiences – finally, producing an attachment-related personality profile. Their assessments led to concentrate attention on the dimensions of ambivalence and depressivity on the one hand (Harry and Erika) and obsessivity and somaticity on the other (Jordan and Beth). Our study focused on these dimensions exclusively for the specific characteristics of the recruited patients. Other patients could have taken the analysis in different directions depending on their prominent clinical features (dissociative symptoms and separation anxiety, for example). While in the case of Harry and Jordan, data appeared coherent – suggesting predominant ambivalence and obsessivity, respectively – some dissonant elements were found in the stories of the other two patients (as summarized in Table 4). Only interpreting the ambiguous items in a non-default way could restore overall consistency. When Erika's ambivalent answers were put in a depressive perspective and Beth's obsessive answers in a somatic one, inconsistencies were solved, and their comprehensive stories became clear.

**(S2b)** The therapists conducted the first 18 months of treatment unaware of their patients' ACQ scores. In other words, treatment was entirely independent of questionnaire results. Since attachment-informed therapy requires time to build a secure relationship (Berry & Danquah, 2016; Bowlby, 1969/1982; Gold, 2011; Slade & Holmes, 2019), this t-period was deemed indispensable to let the clinicians reach a reliable understanding of the cases. Following their patients' inputs, the therapists were gradually able to gather pieces of information concerning their most clinically relevant attachment dimensions and put them together into a coherent narrative. In other words, the clinicians built an informal attachment-related personality profile of their patients, which informed their therapeutic decisions. The ACQ allows the collection of a range of relevant data in a structured way, supporting the clinician in this process and offering potential cues for discussion with the patient to their treatment benefit.

**(S3)** We first tested the blind scoring of the ACQ against the information therapists gathered throughout treatment. To ensure an unbiased test, the clinicians reported their profiles in written form before knowing about their patients' ACQ responses. This information led us to confirm that – while Harry's and Jordan's predominant dimensions were ambivalence and obsessivity (respectively) – Erika and Beth gave non-default ambivalent and obsessive answers. As a depressive, Erika suffered from a pervasive sense of loneliness (Achterbergh et al., 2020; Mann et al., 2022) that involved feelings of rejection and abandonment typical of ambivalence (Collins, 1996; Fraley et al., 2000). Therefore, her depressivity was consistent with high scores on the ambivalent items. Similarly, as a somatic, Beth tended to focus on social acceptance and to feel compelled to comply with social standards (Gagliardi, 2021; Nardi & Bellantuono, 2008; Picardi et al., 2003). This inclination led her to strive for extraordinary achievements and recognition by following rigid rules. However, since the same tendency to adhere to a strict code of conduct is also a typical obsessive manifestation (Dostal & Pilkington, 2023; Mancini et al., 2018), it is not surprising that she scored high on the obsessive items. In this case, it is worth noting that the sense of guilt characterizing Jordan and Beth were also clinically different. While Jordan felt morally obliged to abide by his code of conduct, Beth followed her rules to pursue social recognition and affiliation. These two tendencies corresponded to a sense of deontological and altruistic guilt, respectively (Mancini & Gangemi, 2015).

**(S4)** We concluded our study by testing the blind ACQ scoring against the response explanations directly provided by the patients, which further confirmed the blind interpretations.

While the ACQ provides information that allows the scorer to interpret items and opt for non-default meaning attributions (as illustrated in Table 4), standard personality questionnaires do not support an alternative interpretation. In cases like those presented here, a univocal attribution would entail producing a misleading profile. In particular, Erika and Beth would be mistakenly considered ambivalent and obsessive. Moreover, as evident from our discussion, differences in personality assessments can lead to expecting different clinical features – and, hence, making different treatment choices. In our case, the ambivalent and the depressive do not experience the same sense of rejection.

And the obsessive and somatic experience a sense of guilt essentially different (deontological and altruistic, respectively).

Finally, it is noteworthy that the same rationale applied through the ACQ can be instantiated by relying on another personality inventory and its reference framework. Using the BFI, for example, we will refer to the Big Five to give items alternative interpretations (e.g., 'being talkative' out of openness or 'worrying a lot' out of conscientiousness) (cf. 1.1). However, using standard questionnaires to implement this feature would require generating extra-scale information that they do not include.

In conclusion, our results suggest that the self-report assessment of personality – and the possible consequent clinical decision-making – could be significantly improved by instruments that allow the scorer to discern between the possible different meanings respondents can convey with their answers. Designing such tools requires leaving the current statistical standard procedures – such as Factor Analysis – for more advanced pattern recognizers. By mimicking the human understanding of deep personal meanings, AI could significantly enhance personality assessment, overcoming the limitations of the current standard statistical methods.

## 4.1 Limitations and future work

Our study is preliminary and presents several relevant limitations, which also suggest directions for future research. We discuss here a few of them.

(1) Given the characteristics of our participants, we considered alternative interpretations concerning ambivalence and obsessivity. Other attachment/personality dimensions will need to be investigated. Administering the ACQ to a number of clinical patients allowed us to identify various questionnaire items that were given non-default meanings but could not be covered in this work. Phobic items are sometimes read from a somatic perspective, and depressive items from a disorganized one, for example. As discussed next, collecting more data will allow us to deepen our understanding of the phenomenon.

(2) We based our argument on four case studies and qualitative analyses. This limited sample size does not allow for applying statistical procedures and generalizing results. Therefore, collecting more data is indispensable. With this purpose, we are conducting multiple additional studies. In the first one, we are administering the ACQ to a large clinical sample to gain statistical knowledge about the relevance of interpretation to personality assessment. In another study, we are applying the *think-aloud protocol* to the ACQ – asking subjects to speak out their thoughts while completing it. This procedure will allow us to directly evaluate the different meanings individuals can attribute to the items. Finally, we will use the collected data to develop a machine learning-based (ML) model capable of scoring the ACQ.

Introducing the potential of ML in this study serves two distinct objectives: (a) enhancing personality assessment and profiling by using multiple cues and (b) emulating human comprehension of profound personal experiences, transcending the constraints of conventional statistical methods. Technically, these goals are interrelated and can be pursued through suitable methods. Notably, neural networks and decision trees offer valuable assistance in achieving these objectives. Several existing approaches are already advancing in this direction, such as deep learning-based methods (Başaran & Ejimogu, 2021; El Bahri et al., 2023; Zhao et al., 2022). The only limitation of applying deep learning approaches is their reliance on extensive data. However, we believe a promising starting point could involve using decision trees and neural models that do not depend on large datasets. Furthermore, statistical techniques such as bootstrap and boosting can be employed to address data scarcity. An issue that may arise with deep networks pertains to the "explainability" of model outputs (e.g., estimated attachment dimensions) in relation to their corresponding input data (e.g., questionnaire entries).

Explainable Artificial Intelligence is still a matter of research (Ras et al., 2020). However, this could be a further opportunity to investigate the most relevant questionnaire entries adopted by the ML model for making the decision since the lack of data would orient towards avoiding deep learning approaches. Additionally, since similar entry configurations may be associated with different attachment dimensions, supplementary cues may be required. In other words, we may need to rely on multiple classification approaches, a strategy employed in various domains, including plant disease detection, cancer diagnosis, and biometric recognition (Aburomman & Reaz, 2017; Concas et al., 2022; Fang et al., 2020; Fotouhi et al., 2019; Manavalan, 2020; Micheletto et al., 2022).

(3) Relying on a cognitive-evolutionary approach, we focused on attachment-related aspects of personality using a specific self-report, the ACQ. Nonetheless, the investigation of item interpretation should concern other personality models and inventories, starting from the well-established ones – such as the Big Five model and the instruments assessing its five dimensions (e.g., the NEO-PI or BFI). In the Big Five case, since its inventories were not designed to explore the possible different meanings underpinning each item, information not included in the questionnaires will be necessary. A theoretical framework will also be required to interpret the answers since the model does not refer to any.

## 5. Conclusions

Personality inventories are an invaluable source of data in clinical psychology and psychiatry. Nonetheless, their effectiveness depends on how information is extracted from the collected responses. Traditional methods use consolidated statistical procedures to group items receiving consistent ratings. Despite identifying fundamental dimensional properties, these procedures do not allow for attributing an item alternative interpretations. Relying on our four case studies and the Attachment-Caregiving Questionnaire (ACQ), we showed that different individuals can read the same statements differently, according to their personal meanings. This evidence suggests a self-report designed to account for this possibility could significantly improve personality assessment. Scoring could be realized by a ML model adequately trained, of which we envision the first implementation steps. Our study is preliminary, and further research is indispensable to reach more conclusive results.

## Author contributions

**Author1**: Conceptualization, Data curation, Formal analysis, Investigation, Methodology, Project administration, Resources, Supervision, Validation, Visualization, Writing – original draft, Writing – review & editing; **Author2**: Conceptualization, Data curation, Formal analysis, Investigation, Methodology, Resources, Supervision, Validation, Visualization, Writing – review & editing; **Author3**: Conceptualization, Data curation, Formal analysis, Investigation, Methodology, Resources, Supervision, Validation, Visualization, Writing – review & editing; **Author4**: Conceptualization, Formal analysis, Investigation, Methodology, Supervision, Validation, Visualization, Writing – review & editing.

## Acknowledgments

This study was not funded and not preregistered. The authors want to thank the Institute for Lifecourse Development (ILD), University of Greenwich, London, UK, for the collaboration in producing this work.

**Block001-T006-Q000 - Attachment-Caregiving Questionnaire (ACQ)**

# Attachment-Caregiving Questionnaire (ACQ)

## Attachment-Caregiving Questionnaire (ACQ)

Completing the questionnaire generally takes about an hour. If, for any reason, you want or need to interrupt it, you can resume it later using the same browser and link. The system will retain your answers until the last completed page, even after restarting the device (provided you do not remove the cookies, which maintain the compilation state). A bar at the bottom of the page will indicate your progress.

### Personal Data

This questionnaire is anonymous.
We collect the following data for research purposes only.

Your answers will be identified only by a code of your choice,
which is meant for you to retrieve your profile once it will be ready.

Please, insert the code (at least 7 characters).
It can be made up of your initials followed by a number or a word, for example.

**Code**

## Attachment-Caregiving Questionnaire (ACQ)

### Sex (as assigned at birth)

- ○ Male
- ○ Female
- ○ Other

### Gender

- ○ Male
- ○ Female
- ○ Fluid ○ Trans
- ○ Non-binary
- ○ Queer
- ○ Other

### Sexual Orientation

- ○ Heterosexual
- ○ Homosexual
- ○ Bisexual
- ○ Asexual
- ○ Uncertain
- ○ Other

### Ethnicity

- ○ European
- ○ North African
- ○ Central-West African
- ○ Central-East African
- ○ South African
- ○ Middle-Eastern
- ○ Russian Asian
- ○ Chinese

- Indian
- South-East Asian
- North American
- Central American
- South American
- Australian
- Mixed
- Other

**Age**

**Weight** in Kg

Please, insert your weight in Kg rounding to the nearest ten. For example, insert 65 if your weight is 65,3 Kg or 92 if your weight is 91,7 kg.

**Height** in cm

Please, insert your height in cm – for example, 162 cm or 185 cm.

**Education**

Please, select your highest achievement. (The indicated ages are merely illustrative)

- 01. I didn't attend any school
- 02. I attended school but didn't get any qualification
- 03. I completed elementary school (age 6-10)
- 04. I completed middle school (age 10-13)
- 05. I have a post middle school professional qualification
- 06. I completed high school (age 13-18)

- 07. I have a post high school professional qualification
- 08. I have a bachelor degree
- 09. I have a post-bachelor professional qualification
- 10. I have a post-bachelor specialization
- 11. I have a master degree
- 12. I have a post-master professional qualification
- 13. I have a post-master specialization
- 14. I have a doctorate (PhD)

### Occupation

- 01. Student
- 02. Professional in the Research/University Area (Researcher, University Professor etc.)
- 03. Professional in the Education Area (School Teacher, etc.)
- 04. Professional in the Health/Social Area (Doctor, Psychologist, Educator, Social Worker, etc.)
- 05. Professional in the Technical/Scientific Area (Engineer, Architect, Chemist, Programmer, etc.)
- 06. Professional in the Legal/Administrative Area (Lawyer, Accountant, Bookkeper, Secretary, etc.)
- 07. Professional in the Art/Entertainment Area (Musician, Actor, etc.)
- 08. Professional in the Aesthetic/Wellness Area (Beautician, Coach, etc.)
- 09. Professional in the Religious Area (Vicar, Priest, etc.)
- 10. Professional in the Law Enforcement/Military Area (Policeman, Soldier, etc.)
- 11. Manager of Commercial Activity (Goods)
- 12. Employee of Commercial Activity (Goods) (Shop Assistant, etc.)
- 13. Office Manager (Services)
- 14. Office Clerk (Services) (Customer Care Employee, etc.)
- 15. Manual Worker (Construction Worker, Craftsman, etc.)
- 16. Unemployed
- 17. Other

### Nationality

### Native Language

○ English

○ Other

**Native Language different from English**

**Please, rate your English level**

| No Knowledge | | | | | | | | | | Perfect Knowledge |
|---|---|---|---|---|---|---|---|---|---|---|
| 0 ○ | 1 ○ | 2 ○ | 3 ○ | 4 ○ | 5 ○ | 6 ○ | 7 ○ | 8 ○ | 9 ○ | 10 ○ |

**What is your Native Language:**

**Children**

**Have you ever had children?**

○ Yes

○ No

**How many?**

○ [1] 1 Son

○ [1] 1 Daughter

○ [2] 1 Son and 1 Daughter

○ [2] 2 Sons

○ [2] 2 Daughters

○ [3] 2 Sons and 1 Daughter

○ [3] 2 Daughters and 1 Son

○ [3] 3 Sons

○ [3] 3 Daughters

- [4] 2 Sons and 2 Daughters
- [4] 3 Sons and 1 Daughter
- [4] 3 Daughters and 1 Son
- [4] 4 Sons
- [4] 4 Daughters
- [5] 4 Sons and 1 Daughter
- [5] 4 Daughters and 1 Son
- [5] 3 Sons and 2 Daughters
- [5] 3 Daughters and 2 Sons
- [5] 5 Sons
- [5] 5 Daughters
- [6] Children
- [7] Children
- [8] Children
- [9] Children
- [10] Children
- More than 10 Children

**How old were you when you had your first child?**

**How old were you when you had your last child?**

[ Please, if you only had one child, just select the corresponding item from the list. ]

**Are they all living?**

- Yes
- No

**Siblings**

### Do you have or did you have any siblings?

- Yes
- No

### How many?

- [1] 1 Brother
- [1] 1 Sister
- [2] 1 Brother and 1 Sister
- [2] 2 Brothers
- [2] 2 Sisters
- [3] 2 Brothers and 1 Sister
- [3] 2 Sisters and 1 Brother
- [3] 3 Brothers
- [3] 3 Sisters
- [4] 2 Brothers and 2 Sisters
- [4] 3 Brothers and 1 Sister
- [4] 3 Sisters and 1 Brother
- [4] 4 Brothers
- [4] 4 Sisters
- [5] 4 Brothers and 1 Sister
- [5] 4 Sisters and 1 Brother
- [5] 3 Brothers and 2 Sisters
- [5] 3 Sisters and 2 Brothers
- [5] 5 Brothers
- [5] 5 Sisters
- [6] Siblings
- [7] Siblings
- [8] Siblings
- [9] Siblings
- [10] Siblings
- More than 10 Siblings

**Of your siblings, how old is (or was) the youngest compared to you?**

[ ▾ ]

**Of your siblings, how old is (or was) the oldest compared to you?**

[ Please, if you only have (or had) one sibling, just select the corresponding item from the list. ]

[ ▾ ]

**Are they all living?**

- ○ Yes
- ○ No

**Block003-T017-Q041 - General Condition**

## Attachment-Caregiving Questionnaire (ACQ)

**General condition**

**Psychological Well-being**

**Do you think you currently suffer from any form of psychological discomfort?**

- ○ Yes
- ○ No

**Can you select one of the items on this list, if any, that can describe – at least partially – the core of your discomfort?**

- ○ (1) General existential distress
- ○ (2) Understanding myself better

- (3) Understanding myself and my relationships with others better
- (4) Difficulty reaching my goals
- (5) Self-esteem
- (6) Understanding my own and others' emotions better
- (7) Managing my serious illness
- (8) Managing a serious illness of a loved one
- (9) Managing a loss
- (10) Managing an abuse
- (11) Managing a traumatic event
- (12) Difficulty in dealing with my mother
- (13) Difficulty in dealing with my father
- (14) Difficulty in dealing with both my parents
- (15) Difficulty in dealing with my partner
- (16) Difficulty in dealing with my children
- (17) Difficulty in dealing with several of my family members
- (18) Difficulty in dealing with my friends
- (19) Difficulty in dealing with my colleagues
- (20) Difficulty in my work/study
- (21) Difficulty in some particular social situations
- (22) Difficulty in many social situations, with people in general
- (23) Feeling generally dependent on one or some people
- (24) Feeling driven to behave differently from other people
- (25) Feeling driven to have antisocial behaviors (not accepted by society)
- (26) Feeling driven to have behaviors that are considered evil
- (27) Feeling generally under stress
- (28) Feeling too much anxiety
- (29) Panic attacks
- (30) Concern for my health
- (31) Feeling depressed
- (32) Problems with controlling my impulses
- (33) Problems with food
- (34) Problems with the use of substances
- (35) Problems with gambling
- (36) Obsession with certain fixed ideas
- (37) Trouble living my sexuality
- (38) Distinguishing reality from fantasy

- ○ (39) No item on this list describes it

**Have you ever suffered from panic attacks?**

[ A panic attack is an event of acute fear and physiological activation in which one fears for their health or even life. ]

- ○ I am totally sure I have never suffered from it
- ○ I'm pretty sure I have never suffered from it
- ○ I'm not sure, but I think I have never suffered from it
- ○ I don't know
- ○ I'm not sure, but I think I have suffered from it
- ○ I'm pretty sure I have suffered from it
- ○ I am totally sure I have suffered from it

**Have you ever suffered from depression?**

[ Depression is a period of exceptionally negative mood and thoughts, in which one feels they have no way out, no hope. ]

- ○ I am totally sure I have never suffered from it
- ○ I'm pretty sure I have never suffered from it
- ○ I'm not sure, but I think I have never suffered from it
- ○ I don't know
- ○ I'm not sure, but I think I have suffered from it
- ○ I'm pretty sure I have suffered from it
- ○ I am totally sure I have suffered from it

**Have you ever suffered from an eating disorder (anorexia, bulimia, and/or obesity)?**

[ Anorexia is voluntarily maintaining an insufficient diet which leads to having an extremely low weight (much lower than the norm expected by gender and age). Bulimia consists of having binges (eating a lot in a short time) and trying to compensate for them with subsequent physical activity, laxatives, vomiting, and/or fasting. Obesity is maintaining an extremely excessive weight (far above the norm expected by gender and age). Here, it is understood that these disorders are not caused by physical problems. ]

- I am totally sure I have never suffered from it
- I'm pretty sure I have never suffered from it
- I'm not sure, but I think I have never suffered from it
- I don't know
- I'm not sure, but I think I have suffered from it
- I'm pretty sure I have suffered from it
- I am totally sure I have suffered from it

**Have you ever suffered from an obsessive-compulsive disorder?**

[ An obsessive-compulsive disorder is characterized by obsessions and compulsions. Obsessions are intrusive (namely, that come involuntarily and unwanted) ideas of very unpleasant and disturbing things. Compulsions are acts (physical or mental) that are performed repeatedly (as rituals) to get rid of the aforementioned intrusive ideas. ]

- I am totally sure I have never suffered from it
- I'm pretty sure I have never suffered from it
- I'm not sure, but I think I have never suffered from it
- I don't know
- I'm not sure, but I think I have suffered from it
- I'm pretty sure I have suffered from it
- I am totally sure I have suffered from it

**Have you ever suffered from a post-traumatic stress disorder?**

[ An event is traumatic for us when we perceive it as seriously health-threatening or even lethal – for us or a loved one – and it makes us feel helpless in that situation. A post-traumatic stress disorder is the disturbing and lasting consequence of a traumatic event that cannot be overcome. ]

- I am totally sure I have never suffered from it
- I'm pretty sure I have never suffered from it
- I'm not sure, but I think I have never suffered from it
- I don't know
- I'm not sure, but I think I have suffered from it
- I'm pretty sure I have suffered from it
- I am totally sure I have suffered from it

**Have you ever received a formal diagnosis from a mental health professional?** (if not currently, in the past)

- Yes
- No

**What (main) diagnosis have you received?**

[ We indicate below a partial and simplified list of disorders often diagnosed. If possible, please indicate the one corresponding to the (main) condition you have been diagnosed with. ]

- [1] Depression
- [2] Generalized Anxiety
- [3] Panic
- [4] Agoraphobia
- [5] Social Anxiety
- [6] Separation Anxiety
- [7] Specific Phobia
- [8] Bipolar Disorder
- [9] Anorexia
- [10] Bulimia
- [11] Binge-Eating
- [12] Somatic Disorder (Somatization)
- [13] Body Dysmorphia
- [14] Obsessive-Compulsive Disorder
- [15] Post-traumatic Stress Disorder
- [16] Acute Stress
- [17] Dissociative Identity Disorder
- [18] Amnesia
- [19] Depersonalization/Derealization
- [20] Alcohol-Related Disorder
- [21] Caffeine-Related Disorder
- [22] Cannabis-Related Disorder
- [23] Hallucinogen-Related Disorder
- [24] Opioid-Related Disorder
- [25] Sedative- or Anxiolytic-Related Disorder

- [26] Stimulant-Related Disorder
- [27] Disorder Related to Another Substance
- [28] Gambling-Related Disorder
- [29] Autism
- [30] Asperger Syndrome
- [31] Attention-Deficit/Hyperactivity Disorder (ADHD)
- [32] Specific Learning Disorder
- [33] Motor Disorder (per causa psicologica)
- [34] Psychotic Disorder
- [35] Enuresis
- [36] Encopresis
- [37] Sleep Disorder
- [38] Sexual Dysfunction
- [39] Conduct Disorder
- [40] Impulse-Control Disorder
- [41] Kleptomania
- [42] Paranoid Personality Disorder
- [43] Schizoid Personality Disorder
- [44] Schizotypal Personality Disorder
- [45] Antisocial Personality Disorder (Sociopathy)
- [46] Borderline Personality Disorder
- [47] Histrionic Personality Disorder
- [48] Narcissistic Personality Disorder
- [49] Avoidant Personality Disorder
- [50] Dependent Personality Disorder
- [51] Obsessive-Compulsive Personality Disorder
- [52] Psychopathy
- [53] Pedophilia
- [54] Other Disorder

**Have you been diagnosed with an additional condition besides the main one?**

[ We indicate below a partial and simplified list of disorders often diagnosed. If possible, please indicate the one corresponding to the additional condition you have been diagnosed with – if any. Otherwise, please select 'No Additional Condition'. ]

**Have you ever been helped by (at least) a psychotherapist?**

[ By 'psychotherapist', we mean a mental health professional who supports you – discussing with you – in tackling issues that are problematic for you. ]

- Yes
- No

**Could you please select one of the items on this list, if any, that can describe the principal reason for being helped?**

- (1) General existential distress
- (2) Understanding myself better
- (3) Understanding myself and my relationships with others better
- (4) Difficulty reaching my goals
- (5) Self-esteem
- (6) Understanding my own and others' emotions better
- (7) Managing my serious illness
- (8) Managing a serious illness of a loved one
- (9) Managing a loss
- (10) Managing an abuse
- (11) Managing a traumatic event
- (12) Difficulty in dealing with my mother
- (13) Difficulty in dealing with my father
- (14) Difficulty in dealing with both my parents
- (15) Difficulty in dealing with my partner
- (16) Difficulty in dealing with my children
- (17) Difficulty in dealing with several of my family members
- (18) Difficulty in dealing with my friends
- (19) Difficulty in dealing with my colleagues
- (20) Difficulty in my work/study
- (21) Difficulty in some particular social situations
- (22) Difficulty in many social situations, with people in general
- (23) Feeling generally dependent on one or some people

- (24) Feeling driven to behave differently from other people
- (25) Feeling driven to have antisocial behaviors (not accepted by society)
- (26) Feeling driven to have behaviors that are considered evil
- (27) Feeling generally under stress
- (28) Feeling too much anxiety
- (29) Panic attacks
- (30) Concern for my health
- (31) Feeling depressed
- (32) Problems with controlling my impulses
- (33) Problems with food
- (34) Problems with the use of substances
- (35) Problems with gambling
- (36) Obsession with certain fixed ideas
- (37) Trouble living my sexuality
- (38) Distinguishing reality from fantasy
- (39) No item on this list describes it

**How long have you used psychotherapy altogether?**

- Less than 3 months
- Between 3 and 6 months
- Between 6 months and 1 year
- Between 1 and 2 years
- Between 2 and 3 years
- Between 3 and 4 years
- Between 4 and 5 years
- Between 5 and 6 years
- Between 6 and 7 years
- Between 7 and 8 years
- Between 8 and 9 years
- Between 9 and 10 years
- More than 10 years

**How do you rate the result of this experience?**

- Extremely negative

- Very negative
- Negative
- Neither negative nor positive
- Positive
- Very positive
- Extremely positive

**Has any of these conditions ever been diagnosed to you?**

- (1) Communication Disorder (Language Disorder, Stuttering, etc.)
- (2) Autism/Asperger Syndrom
- (3) Attention-Deficit/Hyperactivity Disorder (ADHD)
- (4) Specific Learning Disorder (SLD) (Dyslexia, Dyscalculia, etc.)
- (5) Tic Disorder
- (6) None of them

**Physical Well-being**

**Do you currently suffer – or think to suffer – from any serious physical-health issue?**

- Yes
- No

**Could you please indicate the main problem?**

[ We indicate below a partial and simplified list of physical issues. If possible, please indicate the one corresponding to your main problem. ]

- — Neoplastic Disease: Breast Cancer
- — Neoplastic Disease: Prostate Cancer
- — Neoplastic Disease: Cancer Different from Previous Ones
- <> Infectious Disease: Hepatitis
- <> Infectious Disease: HIV/AIDS
- <> Infectious Disease: Different from Previous Ones
- — Musculoskeletal Disorder: Trauma-Induced Disability

○ — Musculoskeletal Disorder: Arthritis

○ — Musculoskeletal Disorder: Different from Previous Ones

○ <> Cardiovascular Disorder: Heart Disease

○ <> Cardiovascular Disorder: High Blood Pressure

○ <> Cardiovascular Disorder: Thrombosis

○ <> Cardiovascular Disorder: Different from Previous Ones

○ — Respiratory Disorder: Asthma

○ — Respiratory Disorder: Lung Infection

○ — Respiratory Disorder: Different from Previous Ones

○ <> Digestive Disorder: Crohn's Disease

○ <> Digestive Disorder: Colitis

○ <> Digestive Disorder: Different from Previous Ones

○ — Genitourinary Disorder: Infertility

○ — Genitourinary Disorder: Renal Disease

○ — Genitourinary Disorder: Different from Previous Ones

○ <> Endocrine Disorder: Diabetes

○ <> Endocrine Disorder: Thyroid Disorder

○ <> Endocrine Disorder: Different from Previous Ones

○ — Skin Disorder: Psoriasis

○ — Skin Disorder: Hidradenitis Suppurativa

○ — Skin Disorder: Different from Previous Ones

○ <> Hematological Disorder: Chronic Anemia

○ <> Hematological Disorder: Sickle Cell Disease

○ <> Hematological Disorder: Different from Previous Ones

○ — Immune System Disorder: Lupus

○ — Immune System Disorder: Different from Previous One

○ <> Sense Organs Disorder: Hearing Loss

○ <> Sense Organs Disorder: Vision Loss

○ <> Sense Organs Disorder: Different from Previous Ones

○ — Neurological Disorder: Multiple sclerosis

○ — Neurological Disorder: Muscular dystrophy

○ — Neurological Disorder: Chronic Pain

○ — Neurological Disorder: Cerebral Palsy

○ — Neurological Disorder: Epilepsy

○ — Neurological Disorder: Alzheimer’s disease

○ — Neurological Disorder: Parkinson’s disease

- — Neurological Disorder: Different from Previous Ones
- <> A Problem That Does Not Belong To Any of the Above Categories

## Other Issues

**Is there any other issue – concerning you, other people, or your relationship – that currently seriously worries you?**

- Yes
- No

**Could you please select one of the items on this list, if any, that can describe this issue?**

- — Problems in the relationship with my mother
- — Problems in the relationship with my father
- — Problems in the relationship with both my parents
- — Problems in the relationship with my partner
- — Problems concerning my partner's loyalty
- — Problems concerning my loyalty to my partner
- — Problems in the relationship with my children
- — Problems within my family
- <> Concern for my mother’s situation
- <> Concern for my father’s situation
- <> Concern for my parents' situation
- <> Concern for my partner's situation
- <> Concern for the situation of my child
- <> Concern for the situation of my children
- <> Concern for the situation of other people
- — Problems for the general conditions in the place where I live
- — Problems in my work environment
- <> Unemployment
- <> Fear of losing my job
- <> Immediate financial problem due to job loss
- <> Concern for the course of my business activity
- <> Immediate financial problem due to the failure of my business activity

- ○ <> Problems with family financial management
- ○ — Concern for the future of my family
- ○ — Concern for the future of my Country
- ○ — General insecurity for the future
- ○ <> Problems with justice
- ○ — Another problem

**Current level of stress**

**Overall, in this period, how do you rate your level of stress/concern?**

Period of no stress/concerns — Period of extreme stress/concerns

0 ○ 1 ○ 2 ○ 3 ○ 4 ○ 5 ○ 6 ○ 7 ○ 8 ○ 9 ○ 10 ○

**Block004-T025-Q058 - Specific issues**

## Attachment-Caregiving Questionnaire (ACQ)

**Specific issues**

**Constrictions**

**Are you currently limited or constricted by the presence of someone you care about or their needs?**

- ○ Yes
- ○ No

**Who is this person to you (or who are these people)?**

- ○ – Mother

- – Father
- – Mother and Father
- – Romantic Partner
- – Romantic Partners
- – Child
- – Children
- – Sibling
- – Siblings
- – Other Family Member
- – Other Family Members
- + More than 1 Family Member (from the previous ones)
- – Friend
- – Friends
- – Acquaintance
- – Acquaintances
- – Colleague
- – Colleagues
- – Boss
- – Bosses
- + More than 1 Non-Family (from the previous ones)
- + More People: Family and Non-Family (among all the previous ones)

**Losses**

**Have you recently suffered the loss of someone you cared much about, or do you have reason to believe you might lose them in the near future?**

[ By 'loss' we mean passing away or a definitive separation. ]

- Yes
- No

**Who is this person to you (or who are these people)?**

- – Mother

- – Father
- – Mother and Father
- – Romantic Partner
- – Romantic Partners
- – Child
- – Children
- – Sibling
- – Siblings
- – Other Family Member
- – Other Family Members
- + More than 1 Family Member (from the previous ones)
- – Friend
- – Friends
- – Acquaintance
- – Acquaintances
- – Colleague
- – Colleagues
- – Boss
- – Bosses
- + More than 1 Non-Family (from the previous ones)
- + More People: Family and Non-Family (among all the previous ones)

**Expectations**

**Is there 'anyone' who is currently placing important expectations on you?**

[ This 'anyone' can also be more than one person, or a social group, and even the whole of society. ]

- Yes
- No

**Who is it that makes you feel this pressure?**

- – Mother

- – Father
- – Mother and Father
- – Romantic Partner
- – Romantic Partners
- – Child
- – Children
- – Sibling
- – Siblings
- – Other Family Member
- – Other Family Members
- + More than 1 Family Member (from the previous ones)
- – Friend
- – Friends
- – Acquaintance
- – Acquaintances
- – Colleague
- – Colleagues
- – Boss
- – Bosses
- + More than 1 Non-Family (from the previous ones)
- + More People: Family and Non-Family (among all the previous ones)
- – My Group of Friends
- – Another Social Group
- – Both People and Groups (from all previous ones)
- – Society in General
- – All (People, Groups, Society)

**Care for loved ones**

**Have you ever – for a period of your life – taken care of a loved one who suffered from a serious or even deadly health condition?**

[ This experience may also have occurred to you more than once – for different periods, with different loved ones. ]

- Yes

- ◯ No

**Who was this person to you (or who were these people)?**

- ◯ – Mother
- ◯ – Father
- ◯ – Mother and Father
- ◯ – Romantic Partner
- ◯ – Romantic Partners
- ◯ – Child
- ◯ – Children
- ◯ – Sibling
- ◯ – Siblings
- ◯ – Other Family Member
- ◯ – Other Family Members
- ◯ + More than 1 Family Member (from the previous ones)
- ◯ – Friend
- ◯ – Friends
- ◯ – Acquaintance
- ◯ – Acquaintances
- ◯ – Colleague
- ◯ – Colleagues
- ◯ – Boss
- ◯ – Bosses
- ◯ + More than 1 Non-Family (from the previous ones)
- ◯ + More People: Family and Non-Family (among all the previous ones)

**How long, overall, have you had this experience as a caregiver?**

[ If such an experience is still ongoing, please consider to date. ]

- ◯ No more than 1 day
- ◯ No more than 3 days
- ◯ No more than 1 week
- ◯ No more than 2 weeks
- ◯ No more than 1 month

- No more than 2 months
- No more than 3 months
- No more than 6 months
- No more than 1 year
- No more than 18 months
- No more than 2 years
- No more than 2 years and a half
- No more than 3 years
- No more than 4 years
- No more than 5 years
- No more than 6 years
- No more than 7 years
- No more than 8 years
- No more than 9 years
- No more than 10 years
- More than 10 years

**Deep worry for loved ones**

**Have you ever been – for a period of your life – deeply worried for a loved one who had a serious problem that put them in grave danger without being able to help them?**

[ This is an experience out of the norm (or that should be out of the norm) that may have happened to you more than once – for different periods, with different loved ones. ]

- Yes
- No

**Who was this person to you (or who were these people)?**

- – Mother
- – Father
- – Mother and Father
- – Romantic Partner
- – Romantic Partners

- ○ – Child
- ○ – Children
- ○ – Sibling
- ○ – Siblings
- ○ – Other Family Member
- ○ – Other Family Members
- ○ + More than 1 Family Member (from the previous ones)
- ○ – Friend
- ○ – Friends
- ○ – Acquaintance
- ○ – Acquaintances
- ○ – Colleague
- ○ – Colleagues
- ○ – Boss
- ○ – Bosses
- ○ + More than 1 Non-Family (from the previous ones)
- ○ + More People: Family and Non-Family (among all the previous ones)

**How long, overall, have you had this experience of deep worry?**

[ If such an experience is still ongoing, please consider to date. ]

- ○ No more than 1 day
- ○ No more than 3 days
- ○ No more than 1 week
- ○ No more than 2 weeks
- ○ No more than 1 month
- ○ No more than 2 months
- ○ No more than 3 months
- ○ No more than 6 months
- ○ No more than 1 year
- ○ No more than 18 months
- ○ No more than 2 years
- ○ No more than 2 years and a half
- ○ No more than 3 years
- ○ No more than 4 years

- No more than 5 years
- No more than 6 years
- No more than 7 years
- No more than 8 years
- No more than 9 years
- No more than 10 years
- More than 10 years

**Alcohol, Psychiatric Drugs, and Other Drugs**

**Have you ever used psychoactive substances – in particular, alcohol, psychiatric drugs, or other drugs – that have had (or have) a significant effect on your psychological experience or your general functioning as a person?**

[ A substance is psychoactive when it affects mental processes. ]

- Yes
- No

**Could you please indicate the substance you have used (or are using) that has had (or has) the most significant effect on you or your life?**

[ We indicate below a partial and simplified list of substances. If possible, please indicate the corresponding one or its category. ]

- — Downer: Alcohol
- — Downer: Opioid (e.g. Heroin, Morphine, Methadone, Fentanyl, Percocet)
- — Downer: Cannabis
- — Downer: Percocet (Painkiller)
- — Downer: Other Downer-Substance
- <> Upper: Amphetamine
- <> Upper: Caffeine
- <> Upper: Cocaine
- <> Upper: Nicotine
- <> Upper: Modafinil (Provigil)
- <> Upper: Other Upper-Substance (Stimulant)

- — Psychedelic: Psilocybin
- — Psychedelic: LSD
- — Psychedelic: DMT
- — Psychedelic: Mescaline
- — Psychedelic: Salvia Divinorum
- — Psychedelic: Nitrous Oxide (Laughing Gas)
- — Psychedelic: Scopolamine
- — Psychedelic: Other Psychedelic Substance
- <> Empathogen: MDMA (Ecstasy)
- <> Empathogen: MDA
- <> Empathogen: AMT
- <> Empathogen: Other Empathogen Substance
- — Specific for Psychological Discomfort: Anxiolytic (e.g. Xanax, Valium)
- — Specific for Psychological Discomfort: Antidepressant (e.g. Prozac, Zoloft)
- — Specific for Psychological Discomfort: Mood Stabilizer (e.g. Lithium, Lamotrigine)
- — Specific for Psychological Discomfort: Antipsychotic (e.g. Haldol, Seroquel, Zyprexa)
- <> Substance Not Included in any of the Above Categories

**Do you feel that the effect of this substance is present in the current period?**

- Yes
- No

**Could you please indicate an additional substance you have used (or are using) – if any – that has had (or has) a significant effect on you or your life?**

[ We indicate below a partial and simplified list of substances. If possible, please indicate the corresponding one or its category. Otherwise, please select 'No Additional Substance'. ]

- — Downer: Alcohol
- — Downer: Opioid (e.g. Heroin, Morphine, Methadone, Fentanyl, Percocet)
- — Downer: Cannabis
- — Downer: Percocet (Painkiller)
- — Downer: Other Downer-Substance
- <> Upper: Amphetamine
- <> Upper: Caffeine

- <> Upper: Cocaine
- <> Upper: Nicotine
- <> Upper: Modafinil (Provigil)
- <> Upper: Other Upper-Substance (Stimulant)
- — Psychedelic: Psilocybin
- — Psychedelic: LSD
- — Psychedelic: DMT
- — Psychedelic: Mescaline
- — Psychedelic: Salvia Divinorum
- — Psychedelic: Nitrous Oxide (Laughing Gas)
- — Psychedelic: Scopolamine
- — Psychedelic: Other Psychedelic Substance
- <> Empathogen: MDMA (Ecstasy)
- <> Empathogen: MDA
- <> Empathogen: AMT
- <> Empathogen: Other Empathogen Substance
- — Specific for Psychological Discomfort: Anxiolytic (e.g. Xanax, Valium)
- — Specific for Psychological Discomfort: Antidepressant (e.g. Prozac, Zoloft)
- — Specific for Psychological Discomfort: Mood Stabilizer (e.g. Lithium, Lamotrigine)
- — Specific for Psychological Discomfort: Antipsychotic (e.g. Haldol, Seroquel, Zyprexa)
- <> Substance Not Included in any of the Above Categories
- — No Additional Substance

**Do you feel that the effect of this second substance is present in the current period?**

- Yes
- No

**Block005-T028-Q058 - ACQ-Intro**

## Attachment-Caregiving Questionnaire (ACQ)

This questionnaire consists of 2 main parts: (1) ACQ-CE about your current experience and (2) ACQ-PE about your past experience as a child. Before you start each part, you will be given a description and instructions on how to complete that part.

NB: The questionnaire is completely anonymous: in order for it to be valid, it is essential that answers are given – not only Accurately but also – Authentically, without trying to offer a certain image of oneself or one's family.

**Block006-T034-Q186 - ACQ-CE**

## Attachment-Caregiving Questionnaire (ACQ)

ACQ-CE [Part 1 - Current Experience]

This part is about yourself and your current social experience.
Many items concern your affective relationships, especially with a romantic partner.
In this case, you should consider how you currently feel or would feel in a romantic relationship.
You can think of a current or previous partner if – in the given situation – you would feel like you felt with that partner.
If you never had a romantic relationship, think of a partner and a relationship as you imagine they would be.
Please, concentrate on what you think and how you feel in the given situation.
For each answer, no more than 10 seconds should be sufficient.

Rate every item according to a scale from 0 to 10 as illustrated.
O: The item does not apply to me at all.
10: The item fully applies to me.

When possible, a more appropriate description is indicated,
such as '0: Absolutely False - 10: Absolutely True' or '0: Not Important At All - 10: Extremely Important'.

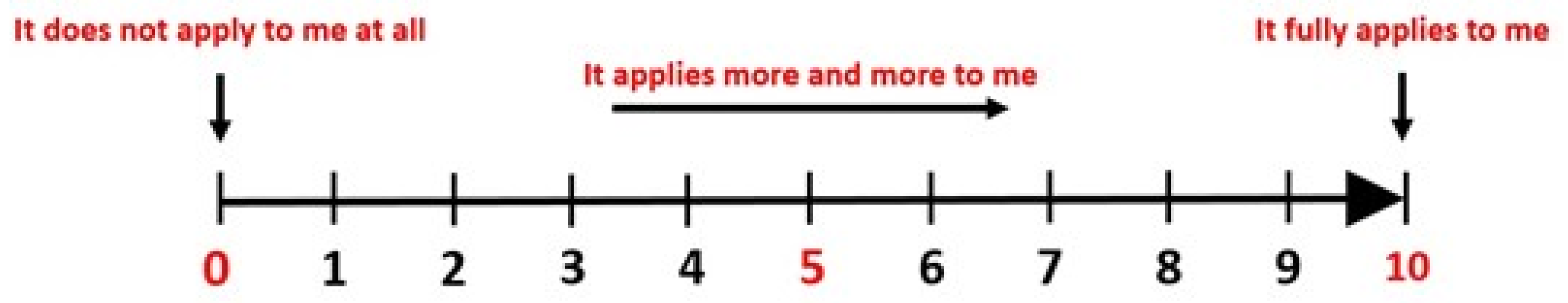

### I currently have a romantic relationship

- Yes
- No

### I have had at least one romantic relationship in my life

- Yes
- No

### I think I have had at least one romantic relationship that has changed my character (positively or negatively)

[ By 'character', we mean above all your way of living a romantic or affective relationship in general. ]

- Yes, at least one relationship that changed my character
- No, no relationship that changed my character

## Attachment-Caregiving Questionnaire (ACQ)

What I currently think and feel:

[ Below, by 'relationship' we mean 'romantic relationship, and we refer to 'my partner' to indicate 'a partner' – who is not necessarily your current partner if you have one ]

### (1) In a relationship, rationality must be the fundamental component

Absolutely Absolutely

False True

0 1 2 3 4 5 6 7 8 9 10

**(2) When one gets emotionally involved, they risk getting trapped in the relationship**

It does not apply to me at all | It fully applies to me

0 1 2 3 4 5 6 7 8 9 10

**(3) In a relationship, I wonder whether my partner really cares about me**

I Never Wonder | I Very Often Wonder

0 1 2 3 4 5 6 7 8 9 10

**(4) In dark times there is never anyone to share your pain with – no matter how much you want it**

It does not apply to me at all | It fully applies to me

0 1 2 3 4 5 6 7 8 9 10

**(5) When it comes to emotions, one needs self-control**

It does not apply to me at all | It fully applies to me

0 1 2 3 4 5 6 7 8 9 10

**(6) For me, it's important that I am always in a safe place or that I can reach one without obstacles in case of emergency**

Not Important | Extremely

At All Important

0 1 2 3 4 5 6 7 8 9 10

**(7) In a relationship, if something goes wrong with me, I allow myself to be consoled**

It does not apply to me at all — It fully applies to me

0 1 2 3 4 5 6 7 8 9 10

**(8) I had a period in which I was overwhelmed by uncontrollable emotions – especially pain and anger – and I felt intolerable sensations**

It does not apply to me at all — It fully applies to me

0 1 2 3 4 5 6 7 8 9 10

**(9) I feel the weight of others' expectations on me**

No, I Don't Feel It At All — Yes, I Feel It Extremely

0 1 2 3 4 5 6 7 8 9 10

**(10) Understanding what others think is essential not to be excluded**

It does not apply to me at all — It fully applies to me

0 1 2 3 4 5 6 7 8 9 10

**(11) I had a period in which I felt so low that I wanted to take my own life**

It does not apply to me at all | | | | | | | | | | It fully applies to me

0 ○ 1 ○ 2 ○ 3 ○ 4 ○ 5 ○ 6 ○ 7 ○ 8 ○ 9 ○ 10 ○

**(12) Sometimes, I feel strongly driven to check I have done everything right to avoid terrible things**

It does not apply to me at all | It fully applies to me

0 ○ 1 ○ 2 ○ 3 ○ 4 ○ 5 ○ 6 ○ 7 ○ 8 ○ 9 ○ 10 ○

**(13) Food and my weight play an important role in my life**

Slightly Important | Absolutely Central

0 ○ 1 ○ 2 ○ 3 ○ 4 ○ 5 ○ 6 ○ 7 ○ 8 ○ 9 ○ 10 ○

**(14) When you leave home to live on your own, it is essential not to go too far – to be always able to get back in case you need help**

It does not apply to me at all | It fully applies to me

0 ○ 1 ○ 2 ○ 3 ○ 4 ○ 5 ○ 6 ○ 7 ○ 8 ○ 9 ○ 10 ○

**(15) In a relationship, I like when my partner shares with me their intimate and profound feelings**

It does not apply to me at all | It fully applies to me

0 ○ 1 ○ 2 ○ 3 ○ 4 ○ 5 ○ 6 ○ 7 ○ 8 ○ 9 ○ 10 ○

**(16) In periods of great stress, I have felt the world around me as somehow unreal**

It Never Happened To Me | | | | | | | | | | It Always Happened To Me

0 ○ 1 ○ 2 ○ 3 ○ 4 ○ 5 ○ 6 ○ 7 ○ 8 ○ 9 ○ 10 ○

**(17) Life requires a strong commitment to facing a destiny of loneliness**

It does not apply to me at all | It fully applies to me

0 ○ 1 ○ 2 ○ 3 ○ 4 ○ 5 ○ 6 ○ 7 ○ 8 ○ 9 ○ 10 ○

**(18) It is useless to hope for words of true comfort when you are down - because nobody will give them to you**

It does not apply to me at all | It fully applies to me

0 ○ 1 ○ 2 ○ 3 ○ 4 ○ 5 ○ 6 ○ 7 ○ 8 ○ 9 ○ 10 ○

**(19) How others see me is important to me**

Not Important At All | Extremely Important

0 ○ 1 ○ 2 ○ 3 ○ 4 ○ 5 ○ 6 ○ 7 ○ 8 ○ 9 ○ 10 ○

**(20) In a relationship, it often seems that my partner is with me only if they have nothing better to do**

It does not apply to me at all | It fully applies to me

0 ○ 1 ○ 2 ○ 3 ○ 4 ○ 5 ○ 6 ○ 7 ○ 8 ○ 9 ○ 10 ○

**(21) I had a period in which I couldn’t feel anything, no emotions at all – as if I were completely empty, although not really sad**

[ Periods of possible 'emotional exhaustion' (burn out) due to the ongoing relationship with people in difficulty – a typical phenomenon of the helping professions, such as doctors, nurses, social workers, therapists, etc. – are to be excluded. ]

It does not apply to me at all | It fully applies to me

0 ○ 1 ○ 2 ○ 3 ○ 4 ○ 5 ○ 6 ○ 7 ○ 8 ○ 9 ○ 10 ○

**(22) In a relationship, I have thoughts about my partner's loyalty**

Never | Very Often

0 ○ 1 ○ 2 ○ 3 ○ 4 ○ 5 ○ 6 ○ 7 ○ 8 ○ 9 ○ 10 ○

**(23) Refraining from taking food can give great satisfaction**

It does not apply to me at all | It fully applies to me

0 ○ 1 ○ 2 ○ 3 ○ 4 ○ 5 ○ 6 ○ 7 ○ 8 ○ 9 ○ 10 ○

**(24) I feel really disgusted by those who don't respect my rules**

It does not apply to me at all | It fully applies to me

0 ○ 1 ○ 2 ○ 3 ○ 4 ○ 5 ○ 6 ○ 7 ○ 8 ○ 9 ○ 10 ○

**(25) Before throwing away certain things, you have to think it over a great deal. Because if you throw them away, you may well be ruined**

Absolutely False | Absolutely True

0 ○ 1 ○ 2 ○ 3 ○ 4 ○ 5 ○ 6 ○ 7 ○ 8 ○ 9 ○ 10 ○

**(26) Being acknowledged by people who count is important to me**

Not Important
At All

Extremely
Important

0 ○ 1 ○ 2 ○ 3 ○ 4 ○ 5 ○ 6 ○ 7 ○ 8 ○ 9 ○ 10 ○

**(27) In any situation, it is important to ensure that you can move freely**

Not Important
At All

Extremely
Important

0 ○ 1 ○ 2 ○ 3 ○ 4 ○ 5 ○ 6 ○ 7 ○ 8 ○ 9 ○ 10 ○

**(28) In some periods of my life, I have felt the anguish of being dirty or contaminated and having to clean up myself**

[ Exceptional periods that involve the entire life context – such as for the spread of a disease – are to be excluded. For example, a period of pandemic (such as that of the corona-virus) is to be excluded. ]

It does not apply
to me at all

It fully applies
to me

0 ○ 1 ○ 2 ○ 3 ○ 4 ○ 5 ○ 6 ○ 7 ○ 8 ○ 9 ○ 10 ○

**(29) When I had a period in which I was overwhelmed by uncontrollable emotions – especially pain and anger – and I felt intolerable sensations, I would have done anything to get out of that state, even hurt myself or directly kill myself**

It does not apply
to me at all

It fully applies
to me

0 ○ 1 ○ 2 ○ 3 ○ 4 ○ 5 ○ 6 ○ 7 ○ 8 ○ 9 ○ 10 ○

**(30) Certain things must be done with the utmost care, my way. Or it bothers me to the point that I feel bad**

It does not apply to me at all | It fully applies to me

0 ○ 1 ○ 2 ○ 3 ○ 4 ○ 5 ○ 6 ○ 7 ○ 8 ○ 9 ○ 10 ○

**(31) At some point, you have to prove to yourself that you can move away from home to explore the world**

Absolutely False | Absolutely True

0 ○ 1 ○ 2 ○ 3 ○ 4 ○ 5 ○ 6 ○ 7 ○ 8 ○ 9 ○ 10 ○

**(32) A strong person doesn’t feel the need to be comforted**

Absolutely False | Absolutely True

0 ○ 1 ○ 2 ○ 3 ○ 4 ○ 5 ○ 6 ○ 7 ○ 8 ○ 9 ○ 10 ○

**(33) It is important to make sure that you don’t get trapped in relationships with people**

Not Important At All | Extremely Important

0 ○ 1 ○ 2 ○ 3 ○ 4 ○ 5 ○ 6 ○ 7 ○ 8 ○ 9 ○ 10 ○

**(34) The slightest doubt that I have done something wrong can make me feel terrible anguish**

It does not apply to me at all | It fully applies to me

0 ○ 1 ○ 2 ○ 3 ○ 4 ○ 5 ○ 6 ○ 7 ○ 8 ○ 9 ○ 10 ○

**(35) In a relationship, if my partner pressurizes me to think as they want, I feel ignored**

It does not apply | It fully applies

to me at all

to me

0 1 2 3 4 5 6 7 8 9 10

**(36) In a relationship, the idea that I can be near my partner makes me feel much more protected**

It does not apply to me at all

It fully applies to me

0 1 2 3 4 5 6 7 8 9 10

**(37) When I have found myself in trouble, I have realized that no one was there to support me with real affection**

It does not apply to me at all

It fully applies to me

0 1 2 3 4 5 6 7 8 9 10

**(38) In a relationship, when I'm feeling down, I keep it to myself and move on**

It does not apply to me at all

It fully applies to me

0 1 2 3 4 5 6 7 8 9 10

**(39) I trust logic much more than emotions**

Absolutely False

Absolutely True

0 1 2 3 4 5 6 7 8 9 10

**(40) You have to be very careful about certain small things. Because, actually, if you make a mistake, you end up getting your life destroyed**

Absolutely

Absolutely

False True

0 1 2 3 4 5 6 7 8 9 10

**(41) In some places – even if absolutely normal – I feel uncomfortable, like I'm constricted or trapped**

It does not apply to me at all / It fully applies to me

0 1 2 3 4 5 6 7 8 9 10

**(42) Failing makes me feel terribly lonely**

It does not apply to me at all / It fully applies to me

0 1 2 3 4 5 6 7 8 9 10

**(43) Doing certain things in front of others makes me (or at least made me) very tense**

It does not apply to me at all / It fully applies to me

0 1 2 3 4 5 6 7 8 9 10

**(44) In a relationship, I think of what I'd do if my partner left me**

I Never Think So / I Very Often Think So

0 1 2 3 4 5 6 7 8 9 10

**(45) Often, if I don't make sure multiple times that I did everything as I should, then the idea can torment me**

It does not apply / It fully applies

to me at all | | | | | | | | | | to me

0 ○ 1 ○ 2 ○ 3 ○ 4 ○ 5 ○ 6 ○ 7 ○ 8 ○ 9 ○ 10 ○

**(46) In a relationship, when I am worried about something, I talk about it with my partner to make myself feel better**

It does not apply to me at all | It fully applies to me

0 ○ 1 ○ 2 ○ 3 ○ 4 ○ 5 ○ 6 ○ 7 ○ 8 ○ 9 ○ 10 ○

**(47) Being disapproved or criticized makes me uncomfortable**

Not Uncomfortable At All | Extremely Uncomfortable

0 ○ 1 ○ 2 ○ 3 ○ 4 ○ 5 ○ 6 ○ 7 ○ 8 ○ 9 ○ 10 ○

**(48) In a relationship, I get angry if I don't get the affection and support I need from my partner**

I Don't Get Angry At All | I Get Extremely Angry

0 ○ 1 ○ 2 ○ 3 ○ 4 ○ 5 ○ 6 ○ 7 ○ 8 ○ 9 ○ 10 ○

**(49) In a relationship, I suffer if I don't feel the affectionate physical touch of my partner**

It does not apply to me at all | It fully applies to me

0 ○ 1 ○ 2 ○ 3 ○ 4 ○ 5 ○ 6 ○ 7 ○ 8 ○ 9 ○ 10 ○

**(50) In a relationship, I know that sooner or later my partner will make me feel terribly bad**

It does not apply | It fully applies

to me at all | to me

0 ○ 1 ○ 2 ○ 3 ○ 4 ○ 5 ○ 6 ○ 7 ○ 8 ○ 9 ○ 10 ○

**(51) Moral issues – what is right or wrong – are at the heart of my thoughts**

It does not apply to me at all | It fully applies to me

0 ○ 1 ○ 2 ○ 3 ○ 4 ○ 5 ○ 6 ○ 7 ○ 8 ○ 9 ○ 10 ○

**(52) I have felt condemned to feel lonely forever**

Absolutely False | Absolutely True

0 ○ 1 ○ 2 ○ 3 ○ 4 ○ 5 ○ 6 ○ 7 ○ 8 ○ 9 ○ 10 ○

**(53) Being left makes me feel like I lost everything**

It does not apply to me at all | It fully applies to me

0 ○ 1 ○ 2 ○ 3 ○ 4 ○ 5 ○ 6 ○ 7 ○ 8 ○ 9 ○ 10 ○

**(54) Thinking of not living up to the expectations on certain occasions makes me very anxious**

It does not apply to me at all | It fully applies to me

0 ○ 1 ○ 2 ○ 3 ○ 4 ○ 5 ○ 6 ○ 7 ○ 8 ○ 9 ○ 10 ○

**(55) Sometimes, thinking of my relationship – irrationally – I felt that I could never leave my partner and, at the same time, I wished I would**

It does not apply | It fully applies

to me at all | to me

0 ◯ 1 ◯ 2 ◯ 3 ◯ 4 ◯ 5 ◯ 6 ◯ 7 ◯ 8 ◯ 9 ◯ 10 ◯

**(56) I feel stuck and constricted when people cross the line I draw for them**

It does not apply to me at all | It fully applies to me

0 ◯ 1 ◯ 2 ◯ 3 ◯ 4 ◯ 5 ◯ 6 ◯ 7 ◯ 8 ◯ 9 ◯ 10 ◯

**(57) For me, it's important to be liked**

Not Important At All | Extremely Important

0 ◯ 1 ◯ 2 ◯ 3 ◯ 4 ◯ 5 ◯ 6 ◯ 7 ◯ 8 ◯ 9 ◯ 10 ◯

**(58) In periods of great stress, I have felt outside of my body**

It Never Happened To Me | It Always Happened To Me

0 ◯ 1 ◯ 2 ◯ 3 ◯ 4 ◯ 5 ◯ 6 ◯ 7 ◯ 8 ◯ 9 ◯ 10 ◯

**(59) Sometimes, I have felt trapped by loved ones who were very close to me, and I have felt the need to feel freer to move**

It does not apply to me at all | It fully applies to me

0 ◯ 1 ◯ 2 ◯ 3 ◯ 4 ◯ 5 ◯ 6 ◯ 7 ◯ 8 ◯ 9 ◯ 10 ◯

**(60) When I had a period in which I felt so low that I wanted to take my own life, I also had thoughts on how to do it concretely**

It does not apply | It fully applies

to me at all

to me

0 1 2 3 4 5 6 7 8 9 10

**(61) For me, it's important to be able to go in and out freely from a situation**

Not Important
At All

Extremely
Important

0 1 2 3 4 5 6 7 8 9 10

**(62) Sometimes, a seemingly small failure makes me feel inexplicably down**

It does not apply
to me at all

It fully applies
to me

0 1 2 3 4 5 6 7 8 9 10

**(63) Generally speaking, I like to feel in my body the strong sensations or emotions given by an exciting substance**

It does not apply
to me at all

It fully applies
to me

0 1 2 3 4 5 6 7 8 9 10

**(64) There is a higher law in the universe - which everyone should respect - and I am extremely careful to respect it**

It does not apply
to me at all

It fully applies
to me

0 1 2 3 4 5 6 7 8 9 10

**(65) Loneliness is the normal condition of life**

It does not apply

It fully applies

to me at all

to me

0 ○ 1 ○ 2 ○ 3 ○ 4 ○ 5 ○ 6 ○ 7 ○ 8 ○ 9 ○ 10 ○

**(66) In some periods of my life, thoughts or images of grave things – happening to others or myself – continuously appeared in my mind without me wanting them to**

It does not apply
to me at all

It fully applies
to me

0 ○ 1 ○ 2 ○ 3 ○ 4 ○ 5 ○ 6 ○ 7 ○ 8 ○ 9 ○ 10 ○

**(67) In a relationship, I never trust to completely put myself in my partner's hands**

It does not apply
to me at all

It fully applies
to me

0 ○ 1 ○ 2 ○ 3 ○ 4 ○ 5 ○ 6 ○ 7 ○ 8 ○ 9 ○ 10 ○

**(68) Strong people keep their suffering to themselves and think about the real problems**

Absolutely
False

Absolutely
True

0 ○ 1 ○ 2 ○ 3 ○ 4 ○ 5 ○ 6 ○ 7 ○ 8 ○ 9 ○ 10 ○

**(69) In a relationship, I desire to discuss my intimate concerns with my partner**

I Never
Desire It

I Always
Desire It

0 ○ 1 ○ 2 ○ 3 ○ 4 ○ 5 ○ 6 ○ 7 ○ 8 ○ 9 ○ 10 ○

**(70) I carefully monitor the internal activation of my body to keep it under control**

I Never
Do It

I Always
Do It

0 ○ 1 ○ 2 ○ 3 ○ 4 ○ 5 ○ 6 ○ 7 ○ 8 ○ 9 ○ 10 ○

**(71) Being in a romantic relationship always leaves me with a sense of fear**

It does not apply to me at all | It fully applies to me

0 ○ 1 ○ 2 ○ 3 ○ 4 ○ 5 ○ 6 ○ 7 ○ 8 ○ 9 ○ 10 ○

**(72) There is something wrong with the very essence of myself**

Absolutely False | Absolutely True

0 ○ 1 ○ 2 ○ 3 ○ 4 ○ 5 ○ 6 ○ 7 ○ 8 ○ 9 ○ 10 ○

**(73) Sensing certain things makes me so disgusted that I feel it on me. And I have to clean myself up as soon as possible**

It does not apply to me at all | It fully applies to me

0 ○ 1 ○ 2 ○ 3 ○ 4 ○ 5 ○ 6 ○ 7 ○ 8 ○ 9 ○ 10 ○

**(74) In periods of great stress, I have felt my body did not really belong to me**

Non Mi È Mai Successo | Mi È Successo Sempre

0 ○ 1 ○ 2 ○ 3 ○ 4 ○ 5 ○ 6 ○ 7 ○ 8 ○ 9 ○ 10 ○

**(75) In a relationship, I'm confident my partner would never leave me**

I am Not Confident At All | I Am Absolutely Confident

0 ○ 1 ○ 2 ○ 3 ○ 4 ○ 5 ○ 6 ○ 7 ○ 8 ○ 9 ○ 10 ○

**(76) In a relationship, crying on the partner's shoulder is for the weak**

Absolutely False | Absolutely True

0 ○ 1 ○ 2 ○ 3 ○ 4 ○ 5 ○ 6 ○ 7 ○ 8 ○ 9 ○ 10 ○

**(77) In a relationship, sometimes, I need to get angry to make my partner hear me**

It does not apply to me at all | It fully applies to me

0 ○ 1 ○ 2 ○ 3 ○ 4 ○ 5 ○ 6 ○ 7 ○ 8 ○ 9 ○ 10 ○

**(78) Only if you fully commit yourself, someone will maybe really love you**

It does not apply to me at all | It fully applies to me

0 ○ 1 ○ 2 ○ 3 ○ 4 ○ 5 ○ 6 ○ 7 ○ 8 ○ 9 ○ 10 ○

**(79) When I don't have the situation under control, I feel constricted, trapped**

It does not apply to me at all | It fully applies to me

0 ○ 1 ○ 2 ○ 3 ○ 4 ○ 5 ○ 6 ○ 7 ○ 8 ○ 9 ○ 10 ○

**(80) In a relationship, it is important to keep one's partner's attention to oneself alive**

Absolutely False | Absolutely True

0 ○ 1 ○ 2 ○ 3 ○ 4 ○ 5 ○ 6 ○ 7 ○ 8 ○ 9 ○ 10 ○

**(81) Always doing the right thing is essential**

Absolutely False | Absolutely True

0 O 1 O 2 O 3 O 4 O 5 O 6 O 7 O 8 O 9 O 10 O

**(82) Generally speaking, I like to feel in my body the strong sensations or emotions given by an exciting activity**

It does not apply to me at all | It fully applies to me

0 O 1 O 2 O 3 O 4 O 5 O 6 O 7 O 8 O 9 O 10 O

**(83) In some periods of my life, I have felt continuously driven to do certain things or have certain thoughts – apparently irrelevant – to avoid terrible consequences**

It does not apply to me at all | It fully applies to me

0 O 1 O 2 O 3 O 4 O 5 O 6 O 7 O 8 O 9 O 10 O

**(84) In important situations, I find it difficult to say no explicitly**

It does not apply to me at all | It fully applies to me

0 O 1 O 2 O 3 O 4 O 5 O 6 O 7 O 8 O 9 O 10 O

**(85) For me, it's important that I can always be easily rescued by a loved one wherever I am**

Not Important At All | Extremely Important

0 O 1 O 2 O 3 O 4 O 5 O 6 O 7 O 8 O 9 O 10 O

**(86) In a relationship, the idea of being left by my partner hardly enters my mind**

| Absolutely False | | | | | | | | | | Absolutely True |
|---|---|---|---|---|---|---|---|---|---|---|
| 0 ○ | 1 ○ | 2 ○ | 3 ○ | 4 ○ | 5 ○ | 6 ○ | 7 ○ | 8 ○ | 9 ○ | 10 ○ |

**(87) The mere memory of those times when I didn't behave as requested makes me relive the embarrassment I felt**

| It does not apply to me at all | | | | | | | | | | It fully applies to me |
|---|---|---|---|---|---|---|---|---|---|---|
| 0 ○ | 1 ○ | 2 ○ | 3 ○ | 4 ○ | 5 ○ | 6 ○ | 7 ○ | 8 ○ | 9 ○ | 10 ○ |

**(88) In a relationship, probably the most positive aspect is the sense of protection that your partner can give you**

| It does not apply to me at all | | | | | | | | | | It fully applies to me |
|---|---|---|---|---|---|---|---|---|---|---|
| 0 ○ | 1 ○ | 2 ○ | 3 ○ | 4 ○ | 5 ○ | 6 ○ | 7 ○ | 8 ○ | 9 ○ | 10 ○ |

**(89) Who loves you the most is also the greatest danger to you**

| Absolutely False | | | | | | | | | | Absolutely True |
|---|---|---|---|---|---|---|---|---|---|---|
| 0 ○ | 1 ○ | 2 ○ | 3 ○ | 4 ○ | 5 ○ | 6 ○ | 7 ○ | 8 ○ | 9 ○ | 10 ○ |

**(90) In a relationship, it is important to know what your partner does when you are not with them**

| Not Important At All | | | | | | | | | | Extremely Important |
|---|---|---|---|---|---|---|---|---|---|---|
| 0 ○ | 1 ○ | 2 ○ | 3 ○ | 4 ○ | 5 ○ | 6 ○ | 7 ○ | 8 ○ | 9 ○ | 10 ○ |

**(91) Not respecting my rules would be unacceptable to me**

It does not apply to me at all | It fully applies to me

0 ◯ 1 ◯ 2 ◯ 3 ◯ 4 ◯ 5 ◯ 6 ◯ 7 ◯ 8 ◯ 9 ◯ 10 ◯

**(92) In a relationship, sometimes, I feel trapped and restricted even if I love my partner**

It does not apply to me at all | It fully applies to me

0 ◯ 1 ◯ 2 ◯ 3 ◯ 4 ◯ 5 ◯ 6 ◯ 7 ◯ 8 ◯ 9 ◯ 10 ◯

**(93) In a relationship, my partner hardly cares about me as much as I care about them**

It does not apply to me at all | It fully applies to me

0 ◯ 1 ◯ 2 ◯ 3 ◯ 4 ◯ 5 ◯ 6 ◯ 7 ◯ 8 ◯ 9 ◯ 10 ◯

**(94) In a relationship, I don't need to be comforted**

It does not apply to me at all | It fully applies to me

0 ◯ 1 ◯ 2 ◯ 3 ◯ 4 ◯ 5 ◯ 6 ◯ 7 ◯ 8 ◯ 9 ◯ 10 ◯

**(95) Rationality is by far more important than emotions**

Absolutely False | Absolutely True

0 ◯ 1 ◯ 2 ◯ 3 ◯ 4 ◯ 5 ◯ 6 ◯ 7 ◯ 8 ◯ 9 ◯ 10 ◯

**(96) In a relationship, sometimes, I think that – if they could – my partner would be with**

**someone else**

It does not apply to me at all | It fully applies to me

0 ◯ 1 ◯ 2 ◯ 3 ◯ 4 ◯ 5 ◯ 6 ◯ 7 ◯ 8 ◯ 9 ◯ 10 ◯

**(97) One needs to be strong and not cry**

It does not apply to me at all | It fully applies to me

0 ◯ 1 ◯ 2 ◯ 3 ◯ 4 ◯ 5 ◯ 6 ◯ 7 ◯ 8 ◯ 9 ◯ 10 ◯

**(98) In a relationship, not receiving the attention I would like to from my partner makes me angry**

Not Angry At All | Extremely Angry

0 ◯ 1 ◯ 2 ◯ 3 ◯ 4 ◯ 5 ◯ 6 ◯ 7 ◯ 8 ◯ 9 ◯ 10 ◯

**(99) In a relationship, I desire to share my intimate and profound feelings with my partner**

It does not apply to me at all | It fully applies to me

0 ◯ 1 ◯ 2 ◯ 3 ◯ 4 ◯ 5 ◯ 6 ◯ 7 ◯ 8 ◯ 9 ◯ 10 ◯

**(100) In my life, I always had to get by by myself**

Absolutely False | Absolutely True

0 ◯ 1 ◯ 2 ◯ 3 ◯ 4 ◯ 5 ◯ 6 ◯ 7 ◯ 8 ◯ 9 ◯ 10 ◯

**(101) Finding real love is just a dream**

It does not apply to me at all | It fully applies to me

0 ◯ 1 ◯ 2 ◯ 3 ◯ 4 ◯ 5 ◯ 6 ◯ 7 ◯ 8 ◯ 9 ◯ 10 ◯

**(102) In a relationship, emotions are only a waste of time**

Absolutely False | Absolutely True

0 ◯ 1 ◯ 2 ◯ 3 ◯ 4 ◯ 5 ◯ 6 ◯ 7 ◯ 8 ◯ 9 ◯ 10 ◯

**(103) In a relationship, my partner somehow makes me feel sure of who I am**

It does not apply to me at all | It fully applies to me

0 ◯ 1 ◯ 2 ◯ 3 ◯ 4 ◯ 5 ◯ 6 ◯ 7 ◯ 8 ◯ 9 ◯ 10 ◯

**(104) Sometimes, the idea that what I did might have terrible consequences becomes an incessant torment that does not give me peace**

It does not apply to me at all | It fully applies to me

0 ◯ 1 ◯ 2 ◯ 3 ◯ 4 ◯ 5 ◯ 6 ◯ 7 ◯ 8 ◯ 9 ◯ 10 ◯

**(105) Being disapproved or criticized makes me feel embarrassed or inadequate**

Not At All | Extremely

0 ◯ 1 ◯ 2 ◯ 3 ◯ 4 ◯ 5 ◯ 6 ◯ 7 ◯ 8 ◯ 9 ◯ 10 ◯

**(106) I think that really reaching someone intimately is impossible**

It does not apply to me at all | It fully applies to me

0○ 1○ 2○ 3○ 4○ 5○ 6○ 7○ 8○ 9○ 10○

**(107) In periods of great stress, I have felt like I was another person, not myself**

It Never Happened To Me | It Always Happened To Me

0○ 1○ 2○ 3○ 4○ 5○ 6○ 7○ 8○ 9○ 10○

**(108) In some periods of my life, thoughts or images of disgusting things continuously appeared in my mind without me wanting them to**

It does not apply to me at all | It fully applies to me

0○ 1○ 2○ 3○ 4○ 5○ 6○ 7○ 8○ 9○ 10○

**(109) In a relationship, I desire the emotional – intimate and profound – support of my partner**

It does not apply to me at all | It fully applies to me

0○ 1○ 2○ 3○ 4○ 5○ 6○ 7○ 8○ 9○ 10○

**(110) In a relationship, I often think that my partner will end up with someone else**

It does not apply to me at all | It fully applies to me

0○ 1○ 2○ 3○ 4○ 5○ 6○ 7○ 8○ 9○ 10○

**(111) In a relationship, I need a partner who hugs and cuddles me**

It does not apply
to me at all

It fully applies
to me

0 ○ 1 ○ 2 ○ 3 ○ 4 ○ 5 ○ 6 ○ 7 ○ 8 ○ 9 ○ 10 ○

**(112) For me, it’s important to feel that others approve of me**

Not Important
At All

Extremely
Important

0 ○ 1 ○ 2 ○ 3 ○ 4 ○ 5 ○ 6 ○ 7 ○ 8 ○ 9 ○ 10 ○

**(113) Sometimes, I think you need to fight to avoid a destiny of loneliness**

It does not apply
to me at all

It fully applies
to me

0 ○ 1 ○ 2 ○ 3 ○ 4 ○ 5 ○ 6 ○ 7 ○ 8 ○ 9 ○ 10 ○

**(114) In periods of great stress, I have felt the world around me as somehow separated from me**

It Never
Happened
To Me

It Always
Happened
To Me

0 ○ 1 ○ 2 ○ 3 ○ 4 ○ 5 ○ 6 ○ 7 ○ 8 ○ 9 ○ 10 ○

**(115) In a relationship, I think my partner prefers others' company to mine**

I Never
Think So

I Very Often
Think So

0 ○ 1 ○ 2 ○ 3 ○ 4 ○ 5 ○ 6 ○ 7 ○ 8 ○ 9 ○ 10 ○

**(116) Not meeting others’ expectations makes me feel inadequate**

It does not apply

It fully applies

to me at all

to me

0 ◯ 1 ◯ 2 ◯ 3 ◯ 4 ◯ 5 ◯ 6 ◯ 7 ◯ 8 ◯ 9 ◯ 10 ◯

**(117) When I had a period in which I couldn't feel anything, no emotions at all – as if I were completely empty, although not really sad – I wanted to die**

[ Periods of possible 'emotional exhaustion' (burn out) due to the ongoing relationship with people in difficulty – a typical phenomenon of the helping professions, such as doctors, nurses, social workers, therapists, etc. – are to be excluded. ]

It does not apply to me at all

It fully applies to me

0 ◯ 1 ◯ 2 ◯ 3 ◯ 4 ◯ 5 ◯ 6 ◯ 7 ◯ 8 ◯ 9 ◯ 10 ◯

**(118) When I get attached to someone, I immediately think I could lose them**

It does not apply to me at all

It fully applies to me

0 ◯ 1 ◯ 2 ◯ 3 ◯ 4 ◯ 5 ◯ 6 ◯ 7 ◯ 8 ◯ 9 ◯ 10 ◯

**(119) In some periods of my life, I have had terrible thoughts that – even if I didn't want to – kept coming to mind and forced me to do something to get rid of them**

It does not apply to me at all

It fully applies to me

0 ◯ 1 ◯ 2 ◯ 3 ◯ 4 ◯ 5 ◯ 6 ◯ 7 ◯ 8 ◯ 9 ◯ 10 ◯

**(120) For me, it's hard to get someone's attention and have some intimate emotional closeness**

It does not apply to me at all

It fully applies to me

0 ◯ 1 ◯ 2 ◯ 3 ◯ 4 ◯ 5 ◯ 6 ◯ 7 ◯ 8 ◯ 9 ◯ 10 ◯

**(121) In a relationship, if my partner pressurizes me to be just like they want, I feel personally violated**

It does not apply to me at all | It fully applies to me

0○ 1○ 2○ 3○ 4○ 5○ 6○ 7○ 8○ 9○ 10 ○

**(122) In periods of great stress, I have felt a familiar place as somehow strange or unknown to me**

It does not apply to me at all | It fully applies to me

0○ 1○ 2○ 3○ 4○ 5○ 6○ 7○ 8○ 9○ 10 ○

**(123) With some people, I would never want to disagree**

It does not apply to me at all | It fully applies to me

0○ 1○ 2○ 3○ 4○ 5○ 6○ 7○ 8○ 9○ 10 ○

**(124) There is an obvious order of things, and I feel extremely uncomfortable when it is not respected**

It does not apply to me at all | It fully applies to me

0○ 1○ 2○ 3○ 4○ 5○ 6○ 7○ 8○ 9○ 10 ○

**(125) When I had a period in which I was overwhelmed by uncontrollable emotions – especially pain and anger – and I felt intolerable sensations, I would have done anything to keep who I loved to myself**

It does not apply to me at all | It fully applies to me

0 ○ 1 ○ 2 ○ 3 ○ 4 ○ 5 ○ 6 ○ 7 ○ 8 ○ 9 ○ 10 ○

**Block007-T037-Q186 - ACQ-PE-Intro**

## Attachment-Caregiving Questionnaire (ACQ)

### ACQ-PE [Part 2 - Past Experience]

This part is about your experience in your family of origin as a child.
There will be 3 sections:

(1) about your family in general;
(2) about your mother (or maternal figure);
(3) about your father (or paternal figure).

**Block008-T044-Q204 - ACQ-PE-Family**

## Attachment-Caregiving Questionnaire (ACQ)

### ACQ-PE - Family [1/3]

(1) This section is about your overall experience in your family of origin as a child.
A family can have multiple configurations, and it is not possible to specifically address each of them here.
Therefore, we will refer to a 'mother' and a 'father' who could be your biological parents or not.
They are intended to be, in general, the 2 principal people - whoever they have been - who took care of you as a child and acted as mother and father. In other words, they could be described as your maternal and paternal figures as a child (regardless of kinship, sex, etc.). The only necessary requirement - for us - for a person to be considered as a maternal or paternal figure is that they started taking care of you before you were 6 years old.
If you had neither a maternal nor paternal figure, please ignore the related section and proceed.

Please, answer for how you remember your experience was – not for what you imagine it was or should have been.
You can rely on memories in the form of images, thoughts, or feelings that you recall while thinking of the past.
For each answer, no more than 10 seconds should be sufficient.

Rate every item according to a scale from 0 to 10 as illustrated.
O: The item does not apply to me at all.
10: The item fully applies to me.

When possible, a more appropriate description is indicated,
such as '0: Absolutely False - 10: Absolutely True' or '0: Not Important At All - 10: Extremely Important'.

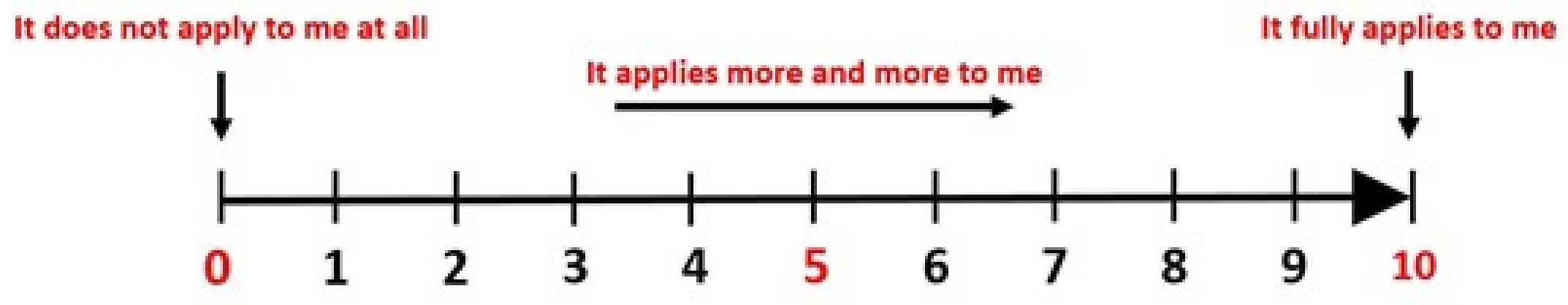

Considering the 2 principal people who took care of you as a child as your maternal and paternal figures:

**What percentage of time did my maternal figure take care – in their own way – of me? (0-100% compared to my paternal figure)**

If you answer "N%", we assume that your paternal figure took care of you – in their own way – for a percentage of time "100-N". For example, if your maternal figure took care of you for 75% of time, we assume that your paternal figure took care of you for 25% of time.

## Attachment-Caregiving Questionnaire (ACQ)

When I was a child, how I remember – in images, thoughts, and feelings – my experience in my family:

**(1) I went outside the home to play with other kids or for other activities not supervised by my parents (none of them)**

[ By parents, we mean your maternal and paternal figure. ]

Never — Very Often

0 ○ 1 ○ 2 ○ 3 ○ 4 ○ 5 ○ 6 ○ 7 ○ 8 ○ 9 ○ 10 ○

**(2) I felt lonely**

Never — Always

0 ○ 1 ○ 2 ○ 3 ○ 4 ○ 5 ○ 6 ○ 7 ○ 8 ○ 9 ○ 10 ○

**(3) I felt I needed help, and nobody helped me**

It does not apply to me at all — It fully applies to me

0 ○ 1 ○ 2 ○ 3 ○ 4 ○ 5 ○ 6 ○ 7 ○ 8 ○ 9 ○ 10 ○

**(4) In my family, sharing certain ideas kept us united**

It does not apply to me at all — It fully applies to me

0 ○ 1 ○ 2 ○ 3 ○ 4 ○ 5 ○ 6 ○ 7 ○ 8 ○ 9 ○ 10 ○

**(5) The family climate was relaxed**

No,
Not Relaxed
At All

Yes,
Completely
Relaxed

0 ◯ 1 ◯ 2 ◯ 3 ◯ 4 ◯ 5 ◯ 6 ◯ 7 ◯ 8 ◯ 9 ◯ 10 ◯

**(6) I had to learn how to get by by myself**

It does not apply
to me at all

It fully applies
to me

0 ◯ 1 ◯ 2 ◯ 3 ◯ 4 ◯ 5 ◯ 6 ◯ 7 ◯ 8 ◯ 9 ◯ 10 ◯

**(7) Meeting family expectations made me feel I belonged to the family**

It does not apply
to me at all

It fully applies
to me

0 ◯ 1 ◯ 2 ◯ 3 ◯ 4 ◯ 5 ◯ 6 ◯ 7 ◯ 8 ◯ 9 ◯ 10 ◯

**(8) I used to take care of at least one member of my family (mother, father, sibling, or other)**

It does not apply
to me at all

It fully applies
to me

0 ◯ 1 ◯ 2 ◯ 3 ◯ 4 ◯ 5 ◯ 6 ◯ 7 ◯ 8 ◯ 9 ◯ 10 ◯

**(9) I felt sad**

Never

Always

0 ◯ 1 ◯ 2 ◯ 3 ◯ 4 ◯ 5 ◯ 6 ◯ 7 ◯ 8 ◯ 9 ◯ 10 ◯

**(10) My parents got along well**

[ By parents, we mean your maternal and paternal figure. If you haven't had one of them, please give any answer – the question will not be considered. ]

| No, Not Well At All | | | | | | | | | | Yes, Completely Well |
|---|---|---|---|---|---|---|---|---|---|---|
| 0 ○ | 1 ○ | 2 ○ | 3 ○ | 4 ○ | 5 ○ | 6 ○ | 7 ○ | 8 ○ | 9 ○ | 10 ○ |

**(11) Initially, leaving home and being left at school made me very nervous and tense: it took me time to stay calmer**

| It does not apply to me at all | | | | | | | | | | It fully applies to me |
|---|---|---|---|---|---|---|---|---|---|---|
| 0 ○ | 1 ○ | 2 ○ | 3 ○ | 4 ○ | 5 ○ | 6 ○ | 7 ○ | 8 ○ | 9 ○ | 10 ○ |

**(12) I hated myself**

| It does not apply to me at all | | | | | | | | | | It fully applies to me |
|---|---|---|---|---|---|---|---|---|---|---|
| 0 ○ | 1 ○ | 2 ○ | 3 ○ | 4 ○ | 5 ○ | 6 ○ | 7 ○ | 8 ○ | 9 ○ | 10 ○ |

**(13) I felt powerless**

| It does not apply to me at all | | | | | | | | | | It fully applies to me |
|---|---|---|---|---|---|---|---|---|---|---|
| 0 ○ | 1 ○ | 2 ○ | 3 ○ | 4 ○ | 5 ○ | 6 ○ | 7 ○ | 8 ○ | 9 ○ | 10 ○ |

**(14) In my family, nobody was expected to have secrets**

| It does not apply to me at all | | | | | | | | | | It fully applies to me |
|---|---|---|---|---|---|---|---|---|---|---|
| 0 ○ | 1 ○ | 2 ○ | 3 ○ | 4 ○ | 5 ○ | 6 ○ | 7 ○ | 8 ○ | 9 ○ | 10 ○ |

**(15) I had to spend much more time at home than most other children**

It does not apply
to me at all

It fully applies
to me

0○ 1○ 2○ 3○ 4○ 5○ 6○ 7○ 8○ 9○ 10○

**(16) My family was united**

No,
Not United
At All

Yes,
Extremely
United

0○ 1○ 2○ 3○ 4○ 5○ 6○ 7○ 8○ 9○ 10○

**(17) My parents could fight quite violently - verbally or physically**

[ By parents, we mean your maternal and paternal figure. If you haven't had one of them, please give any answer – the question will not be considered. ]

It does not apply
to me at all

It fully applies
to me

0○ 1○ 2○ 3○ 4○ 5○ 6○ 7○ 8○ 9○ 10○

**Block009-T050-Q291 - ACQ-PE-Mother**

## Attachment-Caregiving Questionnaire (ACQ)

ACQ-PE - Mother [2/3]

(2) This section is about your experience with your mother (or maternal figure) as a child.
If you did not have a maternal figure who took care of you as a child, please just answer the related question accordingly, ignore this section, and proceed.

Please, answer for how you remember your experience was – not for what you imagine it was or should have been.
You can rely on memories in the form of images, thoughts, or feelings that you recall while thinking of the past.
For each answer, no more than 10 seconds should be sufficient.

Rate every item according to a scale from 0 to 10 as illustrated.
O: The item does not apply to me at all.
10: The item fully applies to me.

When possible, a more appropriate description is indicated,
such as '0: Absolutely False - 10: Absolutely True' or '0: Not Important At All - 10: Extremely Important'.

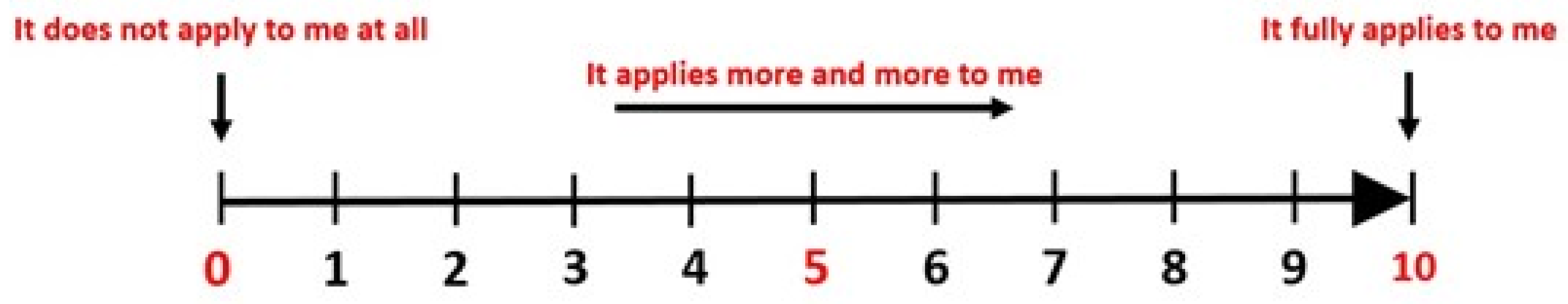

**In my childhood, I had a maternal figure**

- ○ Yes
- ○ No

**My maternal figure – referred to as 'mother' below – has been:**

- ○ My biological mother
- ○ My step-mother
- ○ My grandmother
- ○ My aunt
- ○ My sister
- ○ My cousin (female)
- ○ A family friend (female)
- ○ A nanny (female)
- ○ Another female person

- Another male person
- Another person of non-specified sex

**My mother took care of me since I was:**

[ We can consider a person as a maternal figure only if they started taking care of you before you were 6 years old. In case of a noncontinuous period of time, please consider the earliest part of it. For example, if your mother took care of you when you were between 2 and 4 years old and then between 8 and 13, consider 2-4. ]

**My mother took care of me until I was:**

## Attachment-Caregiving Questionnaire (ACQ)

When I was a child, how I remember – in images, thoughts, and feelings – my experience with my mother:

**(1) My mother blamed me for things I experienced as really anguishing**

It does not apply to me at all — It fully applies to me

0 1 2 3 4 5 6 7 8 9 10

**(2) All mothers are concerned about the health and safety of their children: mine was even more so**

No, Not At All — Yes, Much More So

0 ○ 1 ○ 2 ○ 3 ○ 4 ○ 5 ○ 6 ○ 7 ○ 8 ○ 9 ○ 10 ○

**(3) I feared that my mother would beat me up**

Never | Very Often

0 ○ 1 ○ 2 ○ 3 ○ 4 ○ 5 ○ 6 ○ 7 ○ 8 ○ 9 ○ 10 ○

**(4) I feel anger if I consider that my mother could have thought more about me and my needs**

No, No Anger | Yes, Much Anger

0 ○ 1 ○ 2 ○ 3 ○ 4 ○ 5 ○ 6 ○ 7 ○ 8 ○ 9 ○ 10 ○

**(5) My mother always knew what was appropriate for the situation**

It does not apply to me at all | It fully applies to me

0 ○ 1 ○ 2 ○ 3 ○ 4 ○ 5 ○ 6 ○ 7 ○ 8 ○ 9 ○ 10 ○

**(6) Sometimes, I felt anguish for what my mother might do or say**

It does not apply to me at all | It fully applies to me

0 ○ 1 ○ 2 ○ 3 ○ 4 ○ 5 ○ 6 ○ 7 ○ 8 ○ 9 ○ 10 ○

**(7) I felt bad when I wasn’t or didn’t do as my mother wanted**

It does not apply to me at all | It fully applies to me

0 ○ 1 ○ 2 ○ 3 ○ 4 ○ 5 ○ 6 ○ 7 ○ 8 ○ 9 ○ 10 ○

**(8) My mother could stop talking to me for something I had done**

It does not apply to me at all | It fully applies to me

0 ○ 1 ○ 2 ○ 3 ○ 4 ○ 5 ○ 6 ○ 7 ○ 8 ○ 9 ○ 10 ○

**(9) My mother got mad at me when I did something wrong**

It does not apply to me at all | It fully applies to me

0 ○ 1 ○ 2 ○ 3 ○ 4 ○ 5 ○ 6 ○ 7 ○ 8 ○ 9 ○ 10 ○

**(10) I wished I could spend time with my mother but was rarely able to**

It does not apply to me at all | It fully applies to me

0 ○ 1 ○ 2 ○ 3 ○ 4 ○ 5 ○ 6 ○ 7 ○ 8 ○ 9 ○ 10 ○

**(11) Sometimes, my mother put me under a lot of pressure**

It does not apply to me at all | It fully applies to me

0 ○ 1 ○ 2 ○ 3 ○ 4 ○ 5 ○ 6 ○ 7 ○ 8 ○ 9 ○ 10 ○

**(12) Sometimes, I was worried about what could happen when my mother was with me**

It does not apply to me at all | It fully applies to me

0 ○ 1 ○ 2 ○ 3 ○ 4 ○ 5 ○ 6 ○ 7 ○ 8 ○ 9 ○ 10 ○

**(13) In some situations, the presence of my mother made me feel more self-confident**

No,
Not More
Self-Confident

Yes,
Extremely More
Self-Confident

0 ○ 1 ○ 2 ○ 3 ○ 4 ○ 5 ○ 6 ○ 7 ○ 8 ○ 9 ○ 10 ○

**(14) I refrained from asking comfort from my mother**

It does not apply
to me at all

It fully applies
to me

0 ○ 1 ○ 2 ○ 3 ○ 4 ○ 5 ○ 6 ○ 7 ○ 8 ○ 9 ○ 10 ○

**(15) At some point, I realized that I would never reach my mother's love**

It does not apply
to me at all

It fully applies
to me

0 ○ 1 ○ 2 ○ 3 ○ 4 ○ 5 ○ 6 ○ 7 ○ 8 ○ 9 ○ 10 ○

**(16) My mother was away from home**

Practically
Never

Practically
Always

0 ○ 1 ○ 2 ○ 3 ○ 4 ○ 5 ○ 6 ○ 7 ○ 8 ○ 9 ○ 10 ○

**(17) Letting my mother down was a burden for me**

It does not apply
to me at all

It fully applies
to me

0 ○ 1 ○ 2 ○ 3 ○ 4 ○ 5 ○ 6 ○ 7 ○ 8 ○ 9 ○ 10 ○

**(18) I hugged or kissed my mother to show her how much I loved her**

Never

Very

Often

0 ○ 1 ○ 2 ○ 3 ○ 4 ○ 5 ○ 6 ○ 7 ○ 8 ○ 9 ○ 10 ○

**(19) I remember that sometimes – unfortunately – my mother wasn't there when I needed her**

It does not apply to me at all — It fully applies to me

0 ○ 1 ○ 2 ○ 3 ○ 4 ○ 5 ○ 6 ○ 7 ○ 8 ○ 9 ○ 10 ○

**(20) Sometimes, my mother made me suffer**

It does not apply to me at all — It fully applies to me

0 ○ 1 ○ 2 ○ 3 ○ 4 ○ 5 ○ 6 ○ 7 ○ 8 ○ 9 ○ 10 ○

**(21) I looked for my mother's closeness but I was never able to have it**

It does not apply to me at all — It fully applies to me

0 ○ 1 ○ 2 ○ 3 ○ 4 ○ 5 ○ 6 ○ 7 ○ 8 ○ 9 ○ 10 ○

**(22) Sometimes, my mother kept me waiting too long for her**

It does not apply to me at all — It fully applies to me

0 ○ 1 ○ 2 ○ 3 ○ 4 ○ 5 ○ 6 ○ 7 ○ 8 ○ 9 ○ 10 ○

**(23) My mother expressed disgust at whoever broke her rules**

It does not apply to me at all — It fully applies to me

0 ○ 1 ○ 2 ○ 3 ○ 4 ○ 5 ○ 6 ○ 7 ○ 8 ○ 9 ○ 10 ○

**(24) I could never really know whether my mother was about to blame me for something**

It does not apply to me at all | It fully applies to me

0○ 1○ 2○ 3○ 4○ 5○ 6○ 7○ 8○ 9○ 10 ○

**(25) When I went somewhere, I knew that my mother could always arrive in no time if I needed her**

It does not apply to me at all | It fully applies to me

0○ 1○ 2○ 3○ 4○ 5○ 6○ 7○ 8○ 9○ 10 ○

**(26) My mother had a constant and severe health problem – or so I thought**

It does not apply to me at all | It fully applies to me

0○ 1○ 2○ 3○ 4○ 5○ 6○ 7○ 8○ 9○ 10 ○

**(27) Sometimes, my mother made a fool of me, and I felt humiliated**

It does not apply to me at all | It fully applies to me

0○ 1○ 2○ 3○ 4○ 5○ 6○ 7○ 8○ 9○ 10 ○

**(28) Sometimes, my mother got ferociously angry at me**

It does not apply to me at all | It fully applies to me

0○ 1○ 2○ 3○ 4○ 5○ 6○ 7○ 8○ 9○ 10 ○

**(29) My relationship with my mother was affectionate**

Not Affectionate At All — Very Affectionate

0 ○ 1 ○ 2 ○ 3 ○ 4 ○ 5 ○ 6 ○ 7 ○ 8 ○ 9 ○ 10 ○

**(30) I loved my mother but – thinking of the circumstances with her – I also feel anger**

No, No Anger At All — Yes, Extreme Anger

0 ○ 1 ○ 2 ○ 3 ○ 4 ○ 5 ○ 6 ○ 7 ○ 8 ○ 9 ○ 10 ○

**(31) My mother considered many activities that most children used to do as dangerous**

It does not apply to me at all — It fully applies to me

0 ○ 1 ○ 2 ○ 3 ○ 4 ○ 5 ○ 6 ○ 7 ○ 8 ○ 9 ○ 10 ○

**(32) Sometimes, I had to make an effort to get my mother to notice she should take care of me**

It does not apply to me at all — It fully applies to me

0 ○ 1 ○ 2 ○ 3 ○ 4 ○ 5 ○ 6 ○ 7 ○ 8 ○ 9 ○ 10 ○

**(33) My mother had strict rules and enforced them harshly**

It does not apply to me at all — It fully applies to me

0 ○ 1 ○ 2 ○ 3 ○ 4 ○ 5 ○ 6 ○ 7 ○ 8 ○ 9 ○ 10 ○

**(34) I have some vivid memories of my mother and I who – while playing games – look into each other's eyes and have fun together**

It does not apply to me at all | It fully applies to me

0○ 1○ 2○ 3○ 4○ 5○ 6○ 7○ 8○ 9○ 10○

**(35) Normally, my mother and I thought the same**

It does not apply to me at all | It fully applies to me

0○ 1○ 2○ 3○ 4○ 5○ 6○ 7○ 8○ 9○ 10○

**(36) Sometimes, I had to have a lot of patience with my mother**

It does not apply to me at all | It fully applies to me

0○ 1○ 2○ 3○ 4○ 5○ 6○ 7○ 8○ 9○ 10○

**(37) Sometimes, my mother threatened to kick me out of the house, and I was anguished at the thought**

It does not apply to me at all | It fully applies to me

0○ 1○ 2○ 3○ 4○ 5○ 6○ 7○ 8○ 9○ 10○

**(38) I loved it when my mother hugged and cuddled me**

It does not apply to me at all | It fully applies to me

0○ 1○ 2○ 3○ 4○ 5○ 6○ 7○ 8○ 9○ 10○

**(39) My mother punished me harshly when I did something wrong**

It does not apply to me at all | It fully applies to me

0 ○ 1 ○ 2 ○ 3 ○ 4 ○ 5 ○ 6 ○ 7 ○ 8 ○ 9 ○ 10 ○

**(40) When my mother saw me sad, she asked me affectionately about what happened and tried to console me**

It does not apply to me at all | It fully applies to me

0 ○ 1 ○ 2 ○ 3 ○ 4 ○ 5 ○ 6 ○ 7 ○ 8 ○ 9 ○ 10 ○

**(41) My mother always found I had done something that I shouldn't have done**

It does not apply to me at all | It fully applies to me

0 ○ 1 ○ 2 ○ 3 ○ 4 ○ 5 ○ 6 ○ 7 ○ 8 ○ 9 ○ 10 ○

**(42) My mother and I would both have been in favor if I had been invited to spend 1-2 weeks away from home for an adventurous activity, such as a summer camp, for example**

Not In Favor At All | Extremely In Favor

0 ○ 1 ○ 2 ○ 3 ○ 4 ○ 5 ○ 6 ○ 7 ○ 8 ○ 9 ○ 10 ○

**(43) My mother left home, and I spent the rest of my childhood without her**

It does not apply to me at all | It fully applies to me

0 ○ 1 ○ 2 ○ 3 ○ 4 ○ 5 ○ 6 ○ 7 ○ 8 ○ 9 ○ 10 ○

**(44) When I wasn't sure of something, I asked my mother**

Never | Very Often

0 ○ 1 ○ 2 ○ 3 ○ 4 ○ 5 ○ 6 ○ 7 ○ 8 ○ 9 ○ 10 ○

**(45) I was curious about my mother's tastes and opinions**

Not Curious At All | Extremely Curious

0 ○ 1 ○ 2 ○ 3 ○ 4 ○ 5 ○ 6 ○ 7 ○ 8 ○ 9 ○ 10 ○

**(46) When my mother was at home, I couldn't relax**

It does not apply to me at all | It fully applies to me

0 ○ 1 ○ 2 ○ 3 ○ 4 ○ 5 ○ 6 ○ 7 ○ 8 ○ 9 ○ 10 ○

**(47) I thought that something terrible might happen to my mother**

No, I Never Thought So | Yes, I Very Often Thought So

0 ○ 1 ○ 2 ○ 3 ○ 4 ○ 5 ○ 6 ○ 7 ○ 8 ○ 9 ○ 10 ○

**(48) Sometimes, how things went between me and my mother was quite irritating**

It does not apply to me at all | It fully applies to me

0 ○ 1 ○ 2 ○ 3 ○ 4 ○ 5 ○ 6 ○ 7 ○ 8 ○ 9 ○ 10 ○

**(49) My mother was in need, and I tried to stay close to her**

It does not apply to me at all — It fully applies to me

0 ○ 1 ○ 2 ○ 3 ○ 4 ○ 5 ○ 6 ○ 7 ○ 8 ○ 9 ○ 10 ○

**(50) For many things, I saw my mother as a point of reference - which I liked, or I would have liked, to follow**

It does not apply to me at all — It fully applies to me

0 ○ 1 ○ 2 ○ 3 ○ 4 ○ 5 ○ 6 ○ 7 ○ 8 ○ 9 ○ 10 ○

**(51) Sometimes, I got irritated because I didn't get the attention I needed from my mother**

It does not apply to me at all — It fully applies to me

0 ○ 1 ○ 2 ○ 3 ○ 4 ○ 5 ○ 6 ○ 7 ○ 8 ○ 9 ○ 10 ○

**(52) My mother caressed and hugged me with affection**

Never — Very Often

0 ○ 1 ○ 2 ○ 3 ○ 4 ○ 5 ○ 6 ○ 7 ○ 8 ○ 9 ○ 10 ○

**(53) I could get rather nervous when I had to part with my mother – I remember some of those moments well**

It does not apply to me at all — It fully applies to me

0 ○ 1 ○ 2 ○ 3 ○ 4 ○ 5 ○ 6 ○ 7 ○ 8 ○ 9 ○ 10 ○

**(54) My mother always had some advice to give me**

It does not apply to me at all

It fully applies to me

0 ○ 1 ○ 2 ○ 3 ○ 4 ○ 5 ○ 6 ○ 7 ○ 8 ○ 9 ○ 10 ○

**(55) I remember the warm sound of my mother’s voice and her sweet words when she asked me how I was**

It does not apply to me at all

It fully applies to me

0 ○ 1 ○ 2 ○ 3 ○ 4 ○ 5 ○ 6 ○ 7 ○ 8 ○ 9 ○ 10 ○

**(56) My mother had strict rules that I was always afraid I could fail to respect**

It does not apply to me at all

It fully applies to me

0 ○ 1 ○ 2 ○ 3 ○ 4 ○ 5 ○ 6 ○ 7 ○ 8 ○ 9 ○ 10 ○

**(57) My mother used to follow my activities closely – much more than most other kids' mothers did**

It does not apply to me at all

It fully applies to me

0 ○ 1 ○ 2 ○ 3 ○ 4 ○ 5 ○ 6 ○ 7 ○ 8 ○ 9 ○ 10 ○

**(58) When I needed some comfort, I wanted but couldn’t go to my mother for it**

It does not apply to me at all

It fully applies to me

0 ○ 1 ○ 2 ○ 3 ○ 4 ○ 5 ○ 6 ○ 7 ○ 8 ○ 9 ○ 10 ○

**(59) My mother paid attention to my behavior and blamed me for misbehaving**

It does not apply to me at all

It fully applies to me

0 ○ 1 ○ 2 ○ 3 ○ 4 ○ 5 ○ 6 ○ 7 ○ 8 ○ 9 ○ 10 ○

**(60) Sometimes, my mother's presence did not allow me to feel as free to move as I would have liked**

It does not apply to me at all

It fully applies to me

0 ○ 1 ○ 2 ○ 3 ○ 4 ○ 5 ○ 6 ○ 7 ○ 8 ○ 9 ○ 10 ○

**(61) Following my mother's rules put me under a lot of pressure**

It does not apply to me at all

It fully applies to me

0 ○ 1 ○ 2 ○ 3 ○ 4 ○ 5 ○ 6 ○ 7 ○ 8 ○ 9 ○ 10 ○

**(62) I used to look up to my mother (at least until a certain age)**

It does not apply to me at all

It fully applies to me

0 ○ 1 ○ 2 ○ 3 ○ 4 ○ 5 ○ 6 ○ 7 ○ 8 ○ 9 ○ 10 ○

**(63) Sometimes, my mother insisted on taking care of me, even though I didn't really feel the need for that – I remember some of those moments well**

It does not apply to me at all

It fully applies to me

0 ○ 1 ○ 2 ○ 3 ○ 4 ○ 5 ○ 6 ○ 7 ○ 8 ○ 9 ○ 10 ○

**(64) I thought of my mother and missed her**

Never

Very Often

0 O 1 O 2 O 3 O 4 O 5 O 6 O 7 O 8 O 9 O 10 O

**(65) I longed for my mother's affection, but I was never able to have it**

It does not apply to me at all

It fully applies to me

0 O 1 O 2 O 3 O 4 O 5 O 6 O 7 O 8 O 9 O 10 O

**(66) My mother had a serious problem that could make her leave home for good – or so I thought**

It does not apply to me at all

It fully applies to me

0 O 1 O 2 O 3 O 4 O 5 O 6 O 7 O 8 O 9 O 10 O

**(67) Sometimes, I was preoccupied thinking that my mother wouldn't be there when I needed her**

It does not apply to me at all

It fully applies to me

0 O 1 O 2 O 3 O 4 O 5 O 6 O 7 O 8 O 9 O 10 O

**(68) Sometimes, it seemed like my mother held a grudge against me**

It does not apply to me at all

It fully applies to me

0 O 1 O 2 O 3 O 4 O 5 O 6 O 7 O 8 O 9 O 10 O

**(69) My mother was a kind of dictator**

Absolutely False

Absolutely True

0○ 1○ 2○ 3○ 4○ 5○ 6○ 7○ 8○ 9○ 10○

**(70) I was always worried my mother would take it out on me**

It does not apply to me at all

It fully applies to me

0○ 1○ 2○ 3○ 4○ 5○ 6○ 7○ 8○ 9○ 10 ○

**(71) My mother rarely showed how much she loved me with tenderness and emotion**

Absolutely False

Absolutely True

0○ 1○ 2○ 3○ 4○ 5○ 6○ 7○ 8○ 9○ 10 ○

**(72) I felt the need for my mother’s affection and cuddles**

It does not apply to me at all

It fully applies to me

0○ 1○ 2○ 3○ 4○ 5○ 6○ 7○ 8○ 9○ 10 ○

**(73) I hoped my mother would recognize my qualities or worth**

It does not apply to me at all

It fully applies to me

0○ 1○ 2○ 3○ 4○ 5○ 6○ 7○ 8○ 9○ 10 ○

**(74) In some situations – which seemed normal to most other children – I wasn’t comfortable without my mother's protection**

It does not apply

It fully applies

to me at all

to me

0 ○ 1 ○ 2 ○ 3 ○ 4 ○ 5 ○ 6 ○ 7 ○ 8 ○ 9 ○ 10 ○

**(75) Sometimes, my mother seemed to be mentally far away, like in another world**

It does not apply to me at all

It fully applies to me

0 ○ 1 ○ 2 ○ 3 ○ 4 ○ 5 ○ 6 ○ 7 ○ 8 ○ 9 ○ 10 ○

**(76) Sometimes, I got irritated because my mother interrupted me while I was doing something I liked – I remember some of those moments well**

It does not apply to me at all

It fully applies to me

0 ○ 1 ○ 2 ○ 3 ○ 4 ○ 5 ○ 6 ○ 7 ○ 8 ○ 9 ○ 10 ○

**(77) My mother talked about emotions and feelings such as happiness, sadness, and love**

Never

Very Often

0 ○ 1 ○ 2 ○ 3 ○ 4 ○ 5 ○ 6 ○ 7 ○ 8 ○ 9 ○ 10 ○

**(78) My mother always told or made me understand what was appropriate to do in a situation**

It does not apply to me at all

It fully applies to me

0 ○ 1 ○ 2 ○ 3 ○ 4 ○ 5 ○ 6 ○ 7 ○ 8 ○ 9 ○ 10 ○

**(79) When my mother was at home, she would come and play with me**

Never

Very Often

0 ◯ 1 ◯ 2 ◯ 3 ◯ 4 ◯ 5 ◯ 6 ◯ 7 ◯ 8 ◯ 9 ◯ 10 ◯

**(80) Sometimes, I was scared by my mother**

It does not apply to me at all — It fully applies to me

0 ◯ 1 ◯ 2 ◯ 3 ◯ 4 ◯ 5 ◯ 6 ◯ 7 ◯ 8 ◯ 9 ◯ 10 ◯

**(81) My mother seemed to suffer when I was sad**

No, She Didn't Seem to Suffer — Yes, She Seemed To Suffer Extremely

0 ◯ 1 ◯ 2 ◯ 3 ◯ 4 ◯ 5 ◯ 6 ◯ 7 ◯ 8 ◯ 9 ◯ 10 ◯

**(82) I used to be very close to my mother and maybe I didn't have all the experiences I could have**

It does not apply to me at all — It fully applies to me

0 ◯ 1 ◯ 2 ◯ 3 ◯ 4 ◯ 5 ◯ 6 ◯ 7 ◯ 8 ◯ 9 ◯ 10 ◯

**(83) Sometimes, my mother wanted to know too much about me**

It does not apply to me at all — It fully applies to me

0 ◯ 1 ◯ 2 ◯ 3 ◯ 4 ◯ 5 ◯ 6 ◯ 7 ◯ 8 ◯ 9 ◯ 10 ◯

**Block010-T056-Q378 - ACQ-PE-Father**

## Attachment-Caregiving Questionnaire (ACQ)

ACQ-PE - Father [3/3]

(3) This section is about your experience with your father (or paternal figure) as a child.
If you did not have a paternal figure who took care of you as a child, please just answer the related question accordingly, ignore this section, and proceed.

Please, answer for how you remember your experience was – not for what you imagine it was or should have been.
You can rely on memories in the form of images, thoughts, or feelings that you recall while thinking of the past.
For each answer, no more than 10 seconds should be sufficient.

Rate every item according to a scale from 0 to 10 as illustrated.
O: The item does not apply to me at all.
10: The item fully applies to me.

When possible, a more appropriate description is indicated,
such as '0: Absolutely False - 10: Absolutely True' or '0: Not Important At All - 10: Extremely Important'.

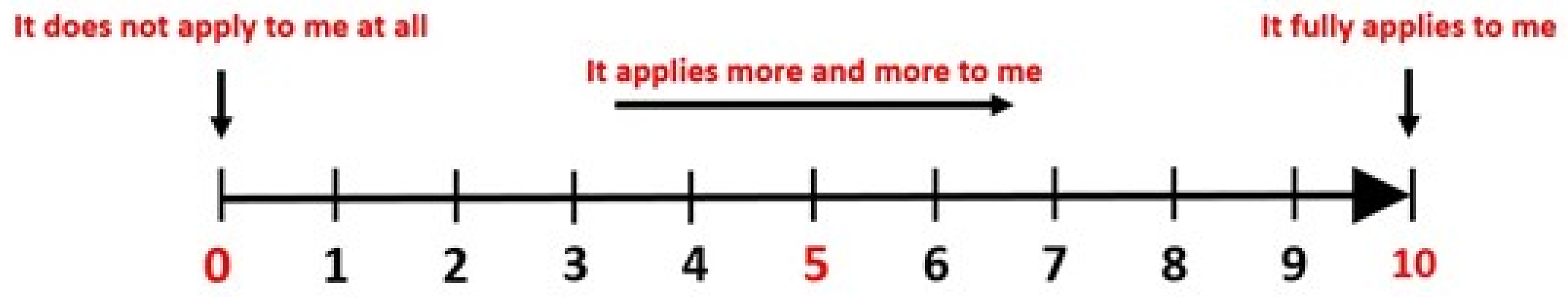

**In my childhood, I had a paternal figure**

- ○ Yes
- ○ No

**My paternal figure - referred to as 'father' below - has been:**

- ○ My biological father
- ○ My step-father

- My grandfather
- My uncle
- My brother
- My cousin (male)
- A family friend (male)
- A nanny (male)
- Another male person
- Another female person
- Another person of non-specified sex

**My father took care of me since I was:**

[ We can consider a person as a paternal figure only if they started taking care of you before you were 6 years old. In case of a noncontinuous period of time, please consider the earliest part of it. For example, if your father took care of you when you were between 2 and 4 years old and then between 8 and 13, consider 2-4. ]

**My father took care of me until I was:**

## Attachment-Caregiving Questionnaire (ACQ)

When I was a child, how I remember – in images, thoughts, and feelings – my experience with my father:

**(1) I felt bad when I wasn't or didn't do as my father wanted**

It does not apply to me at all — It fully applies to me

0 1 2 3 4 5 6 7 8 9 10

**(2) My father expressed disgust at whoever broke his rules**

It does not apply to me at all | It fully applies to me

0○ 1○ 2○ 3○ 4○ 5○ 6○ 7○ 8○ 9○ 10 ○

**(3) My father considered many activities that most children used to do as dangerous**

It does not apply to me at all | It fully applies to me

0○ 1○ 2○ 3○ 4○ 5○ 6○ 7○ 8○ 9○ 10 ○

**(4) I could get rather nervous when I had to part with my father – I remember some of those moments well**

It does not apply to me at all | It fully applies to me

0○ 1○ 2○ 3○ 4○ 5○ 6○ 7○ 8○ 9○ 10 ○

**(5) In some situations, the presence of my father made me feel more self-confident**

No, Not More Self-Confident | Yes, Extremely More Self-Confident

0○ 1○ 2○ 3○ 4○ 5○ 6○ 7○ 8○ 9○ 10 ○

**(6) My father got mad at me when I did something wrong**

It does not apply to me at all | It fully applies to me

0○ 1○ 2○ 3○ 4○ 5○ 6○ 7○ 8○ 9○ 10 ○

**(7) I used to look up to my father (at least until a certain age)**

| It does not apply to me at all | | | | | | | | | | It fully applies to me |
|---|---|---|---|---|---|---|---|---|---|---|
| 0 ○ | 1 ○ | 2 ○ | 3 ○ | 4 ○ | 5 ○ | 6 ○ | 7 ○ | 8 ○ | 9 ○ | 10 ○ |

**(8) Sometimes, my father made me suffer**

| It does not apply to me at all | | | | | | | | | | It fully applies to me |
|---|---|---|---|---|---|---|---|---|---|---|
| 0 ○ | 1 ○ | 2 ○ | 3 ○ | 4 ○ | 5 ○ | 6 ○ | 7 ○ | 8 ○ | 9 ○ | 10 ○ |

**(9) All fathers are concerned about the health and safety of their children: mine was even more so**

| No, Not At All | | | | | | | | | | Yes, Much More So |
|---|---|---|---|---|---|---|---|---|---|---|
| 0 ○ | 1 ○ | 2 ○ | 3 ○ | 4 ○ | 5 ○ | 6 ○ | 7 ○ | 8 ○ | 9 ○ | 10 ○ |

**(10) Following my father’s rules put me under a lot of pressure**

| It does not apply to me at all | | | | | | | | | | It fully applies to me |
|---|---|---|---|---|---|---|---|---|---|---|
| 0 ○ | 1 ○ | 2 ○ | 3 ○ | 4 ○ | 5 ○ | 6 ○ | 7 ○ | 8 ○ | 9 ○ | 10 ○ |

**(11) My father always had some advice to give me**

| It does not apply to me at all | | | | | | | | | | It fully applies to me |
|---|---|---|---|---|---|---|---|---|---|---|
| 0 ○ | 1 ○ | 2 ○ | 3 ○ | 4 ○ | 5 ○ | 6 ○ | 7 ○ | 8 ○ | 9 ○ | 10 ○ |

**(12) I loved my father but – thinking of the circumstances with him – I also feel anger**

| No, No Anger At All | | | | | | | | | | Yes, Extreme Anger |
|---|---|---|---|---|---|---|---|---|---|---|
| 0 ○ | 1 ○ | 2 ○ | 3 ○ | 4 ○ | 5 ○ | 6 ○ | 7 ○ | 8 ○ | 9 ○ | 10 ○ |

**(13) When I needed some comfort, I wanted but couldn’t go to my father for it**

| It does not apply to me at all | | | | | | | | | | It fully applies to me |
|---|---|---|---|---|---|---|---|---|---|---|
| 0 ○ | 1 ○ | 2 ○ | 3 ○ | 4 ○ | 5 ○ | 6 ○ | 7 ○ | 8 ○ | 9 ○ | 10 ○ |

**(14) I refrained from asking comfort from my father**

| It does not apply to me at all | | | | | | | | | | It fully applies to me |
|---|---|---|---|---|---|---|---|---|---|---|
| 0 ○ | 1 ○ | 2 ○ | 3 ○ | 4 ○ | 5 ○ | 6 ○ | 7 ○ | 8 ○ | 9 ○ | 10 ○ |

**(15) Sometimes, I got irritated because I didn’t get the attention I needed from my father**

| It does not apply to me at all | | | | | | | | | | It fully applies to me |
|---|---|---|---|---|---|---|---|---|---|---|
| 0 ○ | 1 ○ | 2 ○ | 3 ○ | 4 ○ | 5 ○ | 6 ○ | 7 ○ | 8 ○ | 9 ○ | 10 ○ |

**(16) I was curious about my father’s tastes and opinions**

| Not Curious At All | | | | | | | | | | Extremely Curious |
|---|---|---|---|---|---|---|---|---|---|---|
| 0 ○ | 1 ○ | 2 ○ | 3 ○ | 4 ○ | 5 ○ | 6 ○ | 7 ○ | 8 ○ | 9 ○ | 10 ○ |

**(17) In some situations – which seemed normal to most other children – I wasn't comfortable without my father's protection**

| It does not apply to me at all | | | | | | | | | | It fully applies to me |
|---|---|---|---|---|---|---|---|---|---|---|
| 0 ○ | 1 ○ | 2 ○ | 3 ○ | 4 ○ | 5 ○ | 6 ○ | 7 ○ | 8 ○ | 9 ○ | 10 ○ |

**(18) My father was a kind of dictator**

| Absolutely False | | | | | | | | | | Absolutely True |
|---|---|---|---|---|---|---|---|---|---|---|
| 0 ○ | 1 ○ | 2 ○ | 3 ○ | 4 ○ | 5 ○ | 6 ○ | 7 ○ | 8 ○ | 9 ○ | 10 ○ |

**(19) I loved it when my father hugged and cuddled me**

| It does not apply to me at all | | | | | | | | | | It fully applies to me |
|---|---|---|---|---|---|---|---|---|---|---|
| 0 ○ | 1 ○ | 2 ○ | 3 ○ | 4 ○ | 5 ○ | 6 ○ | 7 ○ | 8 ○ | 9 ○ | 10 ○ |

**(20) My father paid attention to my behavior and blamed me for misbehaving**

| It does not apply to me at all | | | | | | | | | | It fully applies to me |
|---|---|---|---|---|---|---|---|---|---|---|
| 0 ○ | 1 ○ | 2 ○ | 3 ○ | 4 ○ | 5 ○ | 6 ○ | 7 ○ | 8 ○ | 9 ○ | 10 ○ |

**(21) Sometimes, I was scared by my father**

| It does not apply to me at all | | | | | | | | | | It fully applies to me |
|---|---|---|---|---|---|---|---|---|---|---|
| 0 ○ | 1 ○ | 2 ○ | 3 ○ | 4 ○ | 5 ○ | 6 ○ | 7 ○ | 8 ○ | 9 ○ | 10 ○ |

**(22) Sometimes, my father kept me waiting too long for him**

It does not apply to me at all | It fully applies to me

0○ 1○ 2○ 3○ 4○ 5○ 6○ 7○ 8○ 9○ 10 ○

**(23) My father and I would both have been in favor if I had been invited to spend 1-2 weeks away from home for an adventurous activity, such as a summer camp, for example**

Not In Favor At All | Extremely In Favor

0○ 1○ 2○ 3○ 4○ 5○ 6○ 7○ 8○ 9○ 10 ○

**(24) Sometimes, I was preoccupied thinking that my father wouldn’t be there when I needed him**

It does not apply to me at all | It fully applies to me

0○ 1○ 2○ 3○ 4○ 5○ 6○ 7○ 8○ 9○ 10 ○

**(25) Sometimes, my father got ferociously angry at me**

It does not apply to me at all | It fully applies to me

0○ 1○ 2○ 3○ 4○ 5○ 6○ 7○ 8○ 9○ 10 ○

**(26) My father rarely showed how much he loved me with tenderness and emotion**

Absolutely False | Absolutely True

0○ 1○ 2○ 3○ 4○ 5○ 6○ 7○ 8○ 9○ 10 ○

**(27) My father had a constant and severe health problem – or so I thought**

It does not apply to me at all | It fully applies to me

0○ 1○ 2○ 3○ 4○ 5○ 6○ 7○ 8○ 9○ 10○

**(28) I wished I could spend time with my father but was rarely able to**

It does not apply to me at all | It fully applies to me

0○ 1○ 2○ 3○ 4○ 5○ 6○ 7○ 8○ 9○ 10○

**(29) Sometimes, how things went between me and my father was quite irritating**

It does not apply to me at all | It fully applies to me

0○ 1○ 2○ 3○ 4○ 5○ 6○ 7○ 8○ 9○ 10○

**(30) When I wasn't sure of something, I asked my father**

Never | Very Often

0○ 1○ 2○ 3○ 4○ 5○ 6○ 7○ 8○ 9○ 10○

**(31) Sometimes, I had to make an effort to get my father to notice he should take care of me**

It does not apply to me at all | It fully applies to me

0○ 1○ 2○ 3○ 4○ 5○ 6○ 7○ 8○ 9○ 10○

**(32) I thought that something terrible might happen to my father**

No, I Never Thought So — Yes, I Very Often Thought So

0 ○ 1 ○ 2 ○ 3 ○ 4 ○ 5 ○ 6 ○ 7 ○ 8 ○ 9 ○ 10 ○

**(33) When my father saw me sad, he asked me affectionately about what happened and tried to console me**

It does not apply to me at all — It fully applies to me

0 ○ 1 ○ 2 ○ 3 ○ 4 ○ 5 ○ 6 ○ 7 ○ 8 ○ 9 ○ 10 ○

**(34) Sometimes, it seemed like my father held a grudge against me**

It does not apply to me at all — It fully applies to me

0 ○ 1 ○ 2 ○ 3 ○ 4 ○ 5 ○ 6 ○ 7 ○ 8 ○ 9 ○ 10 ○

**(35) My father had strict rules that I was always afraid I could fail to respect**

It does not apply to me at all — It fully applies to me

0 ○ 1 ○ 2 ○ 3 ○ 4 ○ 5 ○ 6 ○ 7 ○ 8 ○ 9 ○ 10 ○

**(36) Sometimes, my father's presence did not allow me to feel as free to move as I would have liked**

It does not apply to me at all — It fully applies to me

0 ○ 1 ○ 2 ○ 3 ○ 4 ○ 5 ○ 6 ○ 7 ○ 8 ○ 9 ○ 10 ○

**(37) I longed for my father's affection, but I was never able to have it**

It does not apply to me at all | It fully applies to me

0 ○ 1 ○ 2 ○ 3 ○ 4 ○ 5 ○ 6 ○ 7 ○ 8 ○ 9 ○ 10 ○

**(38) My father blamed me for things I experienced as really anguishing**

It does not apply to me at all | It fully applies to me

0 ○ 1 ○ 2 ○ 3 ○ 4 ○ 5 ○ 6 ○ 7 ○ 8 ○ 9 ○ 10 ○

**(39) When my father was at home, I couldn't relax**

It does not apply to me at all | It fully applies to me

0 ○ 1 ○ 2 ○ 3 ○ 4 ○ 5 ○ 6 ○ 7 ○ 8 ○ 9 ○ 10 ○

**(40) Sometimes, my father seemed to be mentally far away, like in another world**

It does not apply to me at all | It fully applies to me

0 ○ 1 ○ 2 ○ 3 ○ 4 ○ 5 ○ 6 ○ 7 ○ 8 ○ 9 ○ 10 ○

**(41) My father always told or made me understand what was appropriate to do in a situation**

It does not apply to me at all | It fully applies to me

0 ○ 1 ○ 2 ○ 3 ○ 4 ○ 5 ○ 6 ○ 7 ○ 8 ○ 9 ○ 10 ○

**(42) Sometimes, I was worried about what could happen when my father was with me**

It does not apply to me at all | It fully applies to me

0 ◯ 1 ◯ 2 ◯ 3 ◯ 4 ◯ 5 ◯ 6 ◯ 7 ◯ 8 ◯ 9 ◯ 10 ◯

**(43) Sometimes, my father insisted on taking care of me, even though I didn't really feel the need for that – I remember some of those moments well**

It does not apply to me at all | It fully applies to me

0 ◯ 1 ◯ 2 ◯ 3 ◯ 4 ◯ 5 ◯ 6 ◯ 7 ◯ 8 ◯ 9 ◯ 10 ◯

**(44) My father left home, and I spent the rest of my childhood without him**

It does not apply to me at all | It fully applies to me

0 ◯ 1 ◯ 2 ◯ 3 ◯ 4 ◯ 5 ◯ 6 ◯ 7 ◯ 8 ◯ 9 ◯ 10 ◯

**(45) Sometimes, I felt anguish for what my father might do or say**

It does not apply to me at all | It fully applies to me

0 ◯ 1 ◯ 2 ◯ 3 ◯ 4 ◯ 5 ◯ 6 ◯ 7 ◯ 8 ◯ 9 ◯ 10 ◯

**(46) My father punished me harshly when I did something wrong**

It does not apply to me at all | It fully applies to me

0 ◯ 1 ◯ 2 ◯ 3 ◯ 4 ◯ 5 ◯ 6 ◯ 7 ◯ 8 ◯ 9 ◯ 10 ◯

**(47) I looked for my father’s closeness but I was never able to have it**

It does not apply | It fully applies

to me at all

to me

0 ○ 1 ○ 2 ○ 3 ○ 4 ○ 5 ○ 6 ○ 7 ○ 8 ○ 9 ○ 10 ○

**(48) Sometimes, my father threatened to kick me out of the house, and I was anguished at the thought**

It does not apply
to me at all

It fully applies
to me

0 ○ 1 ○ 2 ○ 3 ○ 4 ○ 5 ○ 6 ○ 7 ○ 8 ○ 9 ○ 10 ○

**(49) I hoped my father would recognize my qualities or worth**

It does not apply
to me at all

It fully applies
to me

0 ○ 1 ○ 2 ○ 3 ○ 4 ○ 5 ○ 6 ○ 7 ○ 8 ○ 9 ○ 10 ○

**(50) My father could stop talking to me for something I had done**

It does not apply
to me at all

It fully applies
to me

0 ○ 1 ○ 2 ○ 3 ○ 4 ○ 5 ○ 6 ○ 7 ○ 8 ○ 9 ○ 10 ○

**(51) My father was in need, and I tried to stay close to him**

It does not apply
to me at all

It fully applies
to me

0 ○ 1 ○ 2 ○ 3 ○ 4 ○ 5 ○ 6 ○ 7 ○ 8 ○ 9 ○ 10 ○

**(52) I hugged or kissed my father to show him how much I loved him**

Never

Very
Often

0○ 1○ 2○ 3○ 4○ 5○ 6○ 7○ 8○ 9○ 10○

**(53) Letting my father down was a burden for me**

It does not apply to me at all — It fully applies to me

0○ 1○ 2○ 3○ 4○ 5○ 6○ 7○ 8○ 9○ 10○

**(54) For many things, I saw my father as a point of reference - which I liked, or I would have liked, to follow**

It does not apply to me at all — It fully applies to me

0○ 1○ 2○ 3○ 4○ 5○ 6○ 7○ 8○ 9○ 10○

**(55) My father had strict rules and enforced them harshly**

It does not apply to me at all — It fully applies to me

0○ 1○ 2○ 3○ 4○ 5○ 6○ 7○ 8○ 9○ 10○

**(56) Sometimes, my father put me under a lot of pressure**

It does not apply to me at all — It fully applies to me

0○ 1○ 2○ 3○ 4○ 5○ 6○ 7○ 8○ 9○ 10○

**(57) I remember that sometimes – unfortunately – my father wasn't there when I needed him**

It does not apply to me at all — It fully applies to me

0○ 1○ 2○ 3○ 4○ 5○ 6○ 7○ 8○ 9○ 10○

**(58) I felt the need for my father's affection and cuddles**

It does not apply to me at all — It fully applies to me

0○ 1○ 2○ 3○ 4○ 5○ 6○ 7○ 8○ 9○ 10○

**(59) I feel anger if I consider that my father could have thought more about me and my needs**

No, No Anger — Yes, Much Anger

0○ 1○ 2○ 3○ 4○ 5○ 6○ 7○ 8○ 9○ 10○

**(60) I thought of my father and missed him**

Never — Very Often

0○ 1○ 2○ 3○ 4○ 5○ 6○ 7○ 8○ 9○ 10○

**(61) My father used to follow my activities closely – much more than most other kids' fathers did**

It does not apply to me at all — It fully applies to me

0○ 1○ 2○ 3○ 4○ 5○ 6○ 7○ 8○ 9○ 10○

**(62) My father was away from home**

Practically Never — Practically Always

0○ 1○ 2○ 3○ 4○ 5○ 6○ 7○ 8○ 9○ 10○

**(63) I have some vivid memories of my father and I who – while playing games – look into each other's eyes and have fun together**

It does not apply to me at all | It fully applies to me

0○ 1○ 2○ 3○ 4○ 5○ 6○ 7○ 8○ 9○ 10○

**(64) My father had a serious problem that could make him leave home for good – or so I thought**

It does not apply to me at all | It fully applies to me

0○ 1○ 2○ 3○ 4○ 5○ 6○ 7○ 8○ 9○ 10○

**(65) Normally, my father and I thought the same**

It does not apply to me at all | It fully applies to me

0○ 1○ 2○ 3○ 4○ 5○ 6○ 7○ 8○ 9○ 10○

**(66) My father caressed and hugged me with affection**

Never | Very Often

0○ 1○ 2○ 3○ 4○ 5○ 6○ 7○ 8○ 9○ 10○

**(67) I was always worried my father would take it out on me**

It does not apply to me at all | It fully applies to me

0○ 1○ 2○ 3○ 4○ 5○ 6○ 7○ 8○ 9○ 10○

**(68) Sometimes, my father made a fool of me, and I felt humiliated**

It does not apply to me at all | It fully applies to me

0 ◯ 1 ◯ 2 ◯ 3 ◯ 4 ◯ 5 ◯ 6 ◯ 7 ◯ 8 ◯ 9 ◯ 10 ◯

**(69) Sometimes, I had to have a lot of patience with my father**

It does not apply to me at all | It fully applies to me

0 ◯ 1 ◯ 2 ◯ 3 ◯ 4 ◯ 5 ◯ 6 ◯ 7 ◯ 8 ◯ 9 ◯ 10 ◯

**(70) My relationship with my father was affectionate**

Not Affectionate At All | Very Affectionate

0 ◯ 1 ◯ 2 ◯ 3 ◯ 4 ◯ 5 ◯ 6 ◯ 7 ◯ 8 ◯ 9 ◯ 10 ◯

**(71) My father talked about emotions and feelings such as happiness, sadness, and love**

Never | Very Often

0 ◯ 1 ◯ 2 ◯ 3 ◯ 4 ◯ 5 ◯ 6 ◯ 7 ◯ 8 ◯ 9 ◯ 10 ◯

**(72) I could never really know whether my father was about to blame me for something**

It does not apply to me at all | It fully applies to me

0 ◯ 1 ◯ 2 ◯ 3 ◯ 4 ◯ 5 ◯ 6 ◯ 7 ◯ 8 ◯ 9 ◯ 10 ◯

**(73) My father always knew what was appropriate for the situation**

It does not apply to me at all

It fully applies to me

0 ◯ 1 ◯ 2 ◯ 3 ◯ 4 ◯ 5 ◯ 6 ◯ 7 ◯ 8 ◯ 9 ◯ 10 ◯

**(74) Sometimes, I got irritated because my father interrupted me while I was doing something I liked – I remember some of those moments well**

It does not apply to me at all

It fully applies to me

0 ◯ 1 ◯ 2 ◯ 3 ◯ 4 ◯ 5 ◯ 6 ◯ 7 ◯ 8 ◯ 9 ◯ 10 ◯

**(75) When my father was at home, he would come and play with me**

Never

Very Often

0 ◯ 1 ◯ 2 ◯ 3 ◯ 4 ◯ 5 ◯ 6 ◯ 7 ◯ 8 ◯ 9 ◯ 10 ◯

**(76) I was worried that my father would beat me up**

Never

Very Often

0 ◯ 1 ◯ 2 ◯ 3 ◯ 4 ◯ 5 ◯ 6 ◯ 7 ◯ 8 ◯ 9 ◯ 10 ◯

**(77) My father seemed to suffer when I was sad**

No, He Didn't Seem To Suffer

Yes, He Seemed To Suffer Extremely

0 ◯ 1 ◯ 2 ◯ 3 ◯ 4 ◯ 5 ◯ 6 ◯ 7 ◯ 8 ◯ 9 ◯ 10 ◯

**(78) When I went somewhere, I knew that my father could always arrive in no time if I needed him**

It does not apply to me at all — It fully applies to me

0○ 1○ 2○ 3○ 4○ 5○ 6○ 7○ 8○ 9○ 10 ○

**(79) I remember the warm sound of my father's voice and his sweet words when he asked me how I was**

It does not apply to me at all — It fully applies to me

0○ 1○ 2○ 3○ 4○ 5○ 6○ 7○ 8○ 9○ 10 ○

**(80) My father always found I had done something that I shouldn't have done**

It does not apply to me at all — It fully applies to me

0○ 1○ 2○ 3○ 4○ 5○ 6○ 7○ 8○ 9○ 10 ○

**(81) At some point, I realized that I would never reach my father's love**

It does not apply to me at all — It fully applies to me

0○ 1○ 2○ 3○ 4○ 5○ 6○ 7○ 8○ 9○ 10 ○

**(82) Sometimes, my father wanted to know too much about me**

It does not apply to me at all — It fully applies to me

0○ 1○ 2○ 3○ 4○ 5○ 6○ 7○ 8○ 9○ 10 ○

**(83) I used to be very close to my father and maybe I didn't have all the experiences I could have**

It does not apply to me at all — It fully applies to me

0 ○ 1 ○ 2 ○ 3 ○ 4 ○ 5 ○ 6 ○ 7 ○ 8 ○ 9 ○ 10 ○

**Block011-T064-Q392 - Conclusion**

## Attachment-Caregiving Questionnaire (ACQ)

## ACQ - Conclusion

We ask you now if you suffered the loss of your mother or father and conclude with some details on your childhood.

**My mother passed away**

○ Yes
○ No

**When my mother died, my age was:**

[ ▾ ]

**My father passed away**

○ Yes
○ No

**When my father died, my age was:**

## Attachment-Caregiving Questionnaire (ACQ)

**Last questions on your childhood**

[ Additional caregivers and homes ]

**Additional Caregivers**

[ By caregiver, we mean someone who took care of you and was a reference figure for you. ]

**(1) When I was a child/an adolescent, I had an additional caregiver, whom I consider as (or almost as) important as my mother or father**

- ○ Yes
- ○ No

**(2) This caregiver was:**

**(3) This caregiver took care of me since I was:**

**(4) This caregiver took care of me until I was:**

**(5) Overall, this caregiver took care of me for:**

Additional Homes

(1) When I was a child/an adolescent, I lived away from home for a long period/long periods

- Yes
- No

(2) The place where I lived was:

(3) I started living here when I was:

(4) I lived here until I was:

(5) Overall, I lived here for:

Thank you! You can now submit your answers.